\documentclass[twocolumn,english,aps,pra,superscriptaddress]{revtex4-1}
\usepackage{natbib}
\usepackage[T1]{fontenc}
\usepackage[latin9]{inputenc}
\usepackage{babel}
\usepackage{float}
\usepackage{amsmath}
\usepackage{amssymb}
\usepackage{amsfonts}
\usepackage{xcolor}
\usepackage{bbm}
\usepackage[caption=false,font=normalsize,labelfont=sf,textfont=sf]{subfig} 
\usepackage{graphicx,epstopdf}
\usepackage{url}
\usepackage{textcase}
\usepackage{bm}
\usepackage[unicode=true]{hyperref}
\usepackage{listings}
\makeatletter

\newcommand{\pt}{\mathcal{PT}}
\newcommand{\p}{\mathcal{P}}
\newcommand{\T}{\mathcal{T}}
\newcommand{\K}{\mathcal{K}}
\newcommand{\ket}[1]{\vert#1\rangle}
\newcommand{\bra}[1]{\langle#1\vert}

\begin{document}
\title{Parity-time symmetric systems with memory}
\author{Zachary A. Cochran}
\affiliation{Department of Physics, Indiana University - Purdue University Indianapolis (IUPUI), Indianapolis, Indiana 46202 USA}
\author{Avadh Saxena}
\affiliation{Theoretical Division, Los Alamos National Laboratory, Los Alamos, New Mexico 87545 USA}
\author{Yogesh N. Joglekar}
\affiliation{Department of Physics, Indiana University - Purdue University Indianapolis (IUPUI), Indianapolis, Indiana 46202 USA}

\date{\today}
\begin{abstract}
	Classical open systems with balanced gain and loss, i.e. parity-time ($\mathcal{PT}$) symmetric systems, have attracted tremendous attention over the past decade. Their exotic properties arise from exceptional point (EP) degeneracies of non-Hermitian Hamiltonians that govern their dynamics. In recent years, increasingly sophisticated models of $\mathcal{PT}$-symmetric systems with time-periodic (Floquet) driving, time-periodic gain and loss, and time-delayed coupling have been investigated, and such systems have been realized across numerous platforms comprising optics, acoustics, mechanical oscillators, optomechanics, and electrical circuits. Here, we introduce a $\mathcal{PT}$-symmetric (balanced gain and loss) system with memory, and investigate its dynamics analytically and numerically. Our model consists of two coupled $LC$ oscillators with positive and negative resistance, respectively. We introduce memory by replacing either the resistor with a memristor, or the coupling inductor with a meminductor, and investigate the circuit energy dynamics as characterized by $\mathcal{PT}$-symmetric or $\mathcal{PT}$-symmetry broken phases. Due to the resulting nonlinearity, we find that energy dynamics depend on the sign and strength of initial voltages and currents, as well as the distribution of initial circuit energy across its different components.  Surprisingly, at strong inputs, the system exhibits self-organized Floquet dynamics, including $\pt$-symmetry broken phase at vanishingly small dissipation strength. Our results indicate that $\mathcal{PT}$-symmetric systems with memory show a rich landscape.
\end{abstract}

\maketitle

\section{\label{sec:Introduction}Introduction}

Over the past decade, open systems with balanced, spatially separated gain and loss have become a rich area of research. They are described by a special class of non-Hermitian Hamiltonians that are invariant under space- and time-reflections, i.e. parity-time ($\pt$) symmetric Hamiltonians~\cite{Bender2007}. More than two decades ago, Bender and coworkers first introduced a broad class of such continuum Hamiltonians on an infinite line~\cite{Bender1998}, and showed that, in spite of their non-Hermitian nature, they have purely real spectra when the non-Hermiticity is small. The initial, theoretical studies of $\pt$-symmetric Hamiltonians were focused on developing a complex extension of quantum theory~\cite{Bender2002,Mostafazadeh2002,Mostafazadeh2010}. Over the past decade, however, it has become clear that such Hamiltonians describe the dynamics of classical energy density within different parts of a system, in the presence of localized sources or sinks~\cite{Joglekar2013,Longhi2017,LFeng2017,ElGanainy2018}. A non-Hermitian Hamiltonian $H(\gamma)$ is called $\pt$ symmetric if it commutes with an antilinear operator $\pt$, where $\p$ is the parity operator satisfying $\p^2={\mathbbm 1}$, and $\T=U\K$ where $U$ is a unitary operator and $\K$ denotes the (antilinear) complex conjugation operation. At $\gamma=0$, $H$ is Hermitian, has real eigenvalues and a complete set of orthonormal eigenvectors that lead to a unitary time evolution. At small non-Hermiticity $\gamma<\gamma_\mathrm{PT}$, the spectrum of $H(\gamma)$ is purely real, but the non-orthogonal eigenvectors generate a non-unitary, bounded, oscillatory-in-time dynamics. At a critical gain-loss strength $\gamma=\gamma_\mathrm{PT}$, eigenvalues of $H$ become degenerate, as do the corresponding eigenvectors. Such degeneracies, where the non-orthogonal eigenvectors of $H(\gamma_\mathrm{PT})$ do not span the space, are called exceptional points (EPs)~\cite{Kato1995,miri2019,Ozdemir2019}. Beyond the EP, eigenvalues of $H(\gamma)$ occur in complex conjugate pairs.  Due to the antilinear nature of $\pt$ operator that commutes with $H$, an eigenstate $\ket{\psi_\alpha}$ of $H$ with eigenvalue $\epsilon_\alpha$ is a simultaneous eigenstate of the $\pt$ operator with eigenvalue $+1$ if and only if $\epsilon_\alpha$ is real; when $\epsilon_\alpha$ is complex, $\pt\ket{\psi_\alpha}$ gives rise to the eigenvector corresponding to the complex conjugate eigenvalue $\K\epsilon_\alpha=\epsilon^*_\alpha$. 

The transition from a $\pt$-symmetric region (real spectrum) to the $\pt$-symmetry broken region (complex conjugate spectrum) across the EP has been extensively studied in classical wave systems where both gain and loss are readily implemented. Realizations include coupled optical waveguides~\cite{Ruter2010}, fiber loops~\cite{Regen2012}, microring resonators~\cite{Peng2014}, acoustic setups~\cite{Zhu2014}, coupled mechanical oscillators~\cite{Bender2013}, and coupled electrical circuits~\cite{Schindler2011,Wang2020}. Due to quantum fluctuations associated with a linear gain~\cite{Caves1982}, balanced gain and loss configurations are not possible at a quantum level~\cite{Scheel2018}. However, EP degeneracies are also present in dissipative systems with mode-selective losses, thus extending the ideas of $\pt$-symmetry into the quantum domain, where the passive, $\pt$-symmetric systems~\cite{Guo2009,Joglekar2018} have been realized with tabletop~\cite{Klauck2019} and integrated quantum photonics, ultracold atoms~\cite{Li2019a}, a single NV center in diamond~\cite{Wu2019}, and a single superconducting qubit~\cite{Naghiloo2019}. Most of these systems are modeled with a static, $\pt$-symmetric Hamiltonian whose eigenvalues and eigenvectors determine the $\pt$-phase diagram of the system. This landscape is dramatically transformed when one considers $\pt$-symmetric Hamiltonians that are periodic in time with period $T$~\cite{Joglekar2014,Lee2015}. In this case, the eigenvalues $\epsilon_\alpha(t)$ of the instantaneous Hamiltonian do not govern the system dynamics; instead, the $\pt$-phase diagram is determined by an equivalent static Hamiltonian $H_F$ called the Floquet Hamiltonian~\cite{Li2019a,Chitsazi2017,LeonMontiel2018}. Another level of complexity is added when we consider systems described by a nonlinear, $\pt$-symmetric Schrodinger equation~\cite{Konotop2016}. However, in almost all cases~\cite{Wilkey2019} the dynamics are Markovian, i.e. the state of the system at the next instance depends only on its state at present, but not on its history. 

Here, we introduce $\pt$-symmetric systems with memory. Formally, this is to be achieved by making either the gain-loss or the Hermitian part of the Hamiltonian dependent on the history of the system. We use coupled, active and lossy $LC$ circuits as the model~\cite{Schindler2011,Wang2020,Chitsazi2017}, because resistive and inductive elements with memory, i.e. memristors and meminductors, are well understood. The plan of the paper is as follows. In Sec.~\ref{sec:ptcircuits} we review properties of a lossy $RLC$ circuit inductively coupled to an $LC$ circuit with negative resistance $-R$. Modeling and numerical results for a system where the resistance is replaced by a memristor (and the negative resistance is matched) are presented in Sec.~\ref{sec:ptmemristor} and Sec.~\ref{sec:floquet}. Section~\ref{sec:ptmeminductor} has corresponding results for a system where the coupling between the two $LC$ circuits is mediated by a  meminductor instead of a regular inductor. We conclude with discussion in Sec.~\ref{sec:disc}. 


\begin{figure*}
\includegraphics[width=0.22\textwidth]{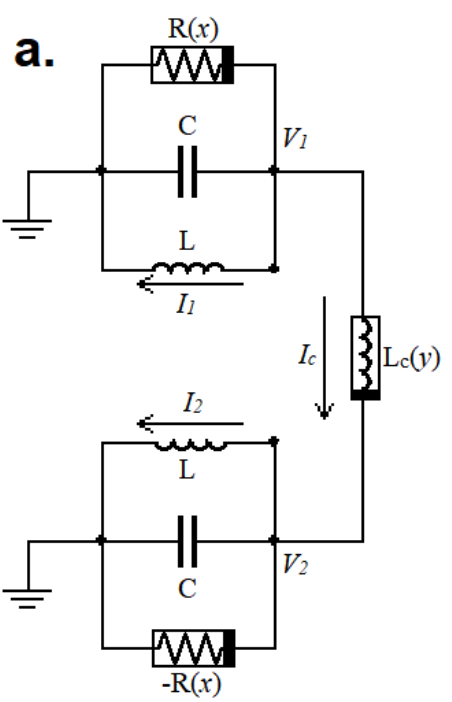}
\includegraphics[width=0.45\textwidth]{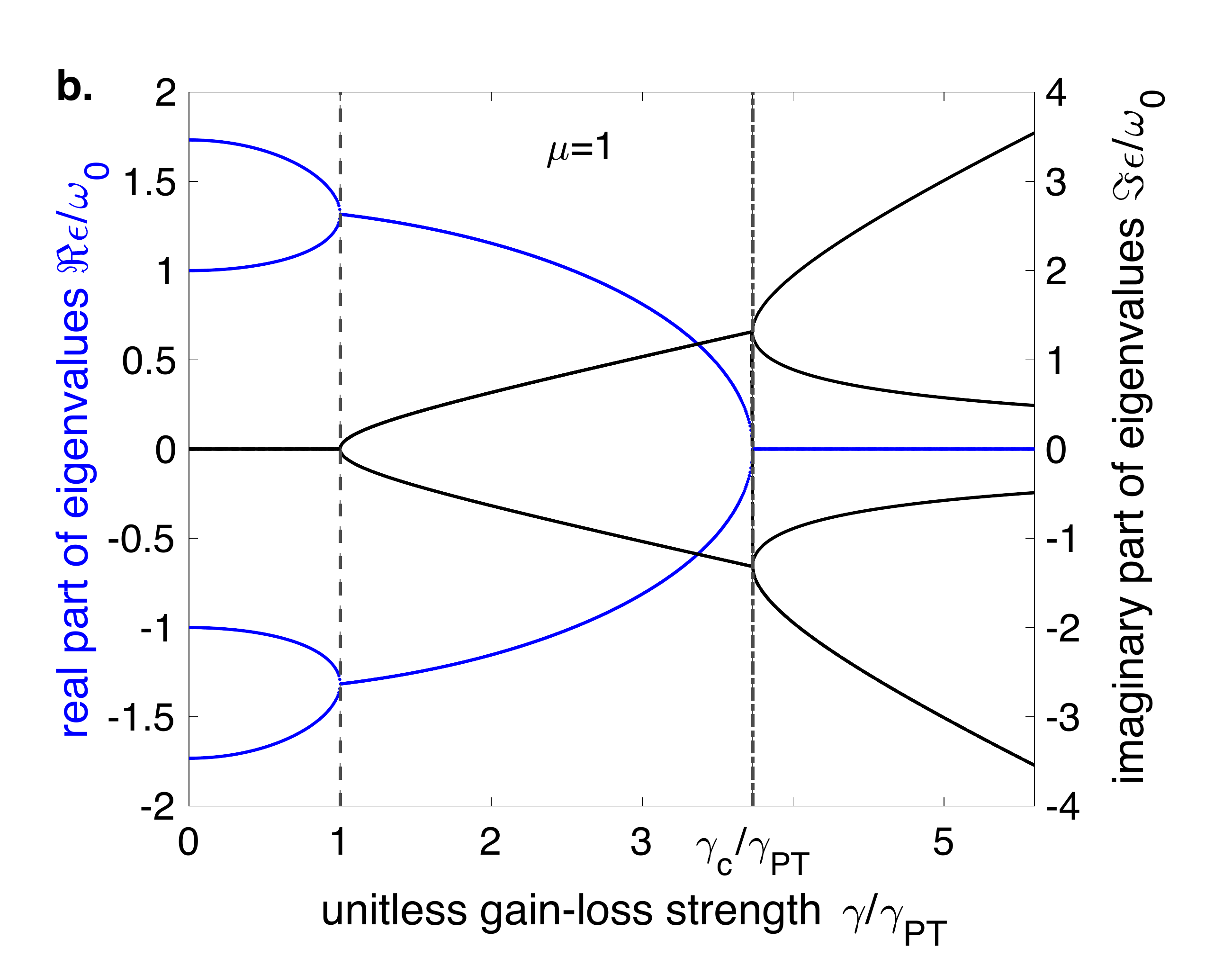}
\includegraphics[width=0.26\textwidth]{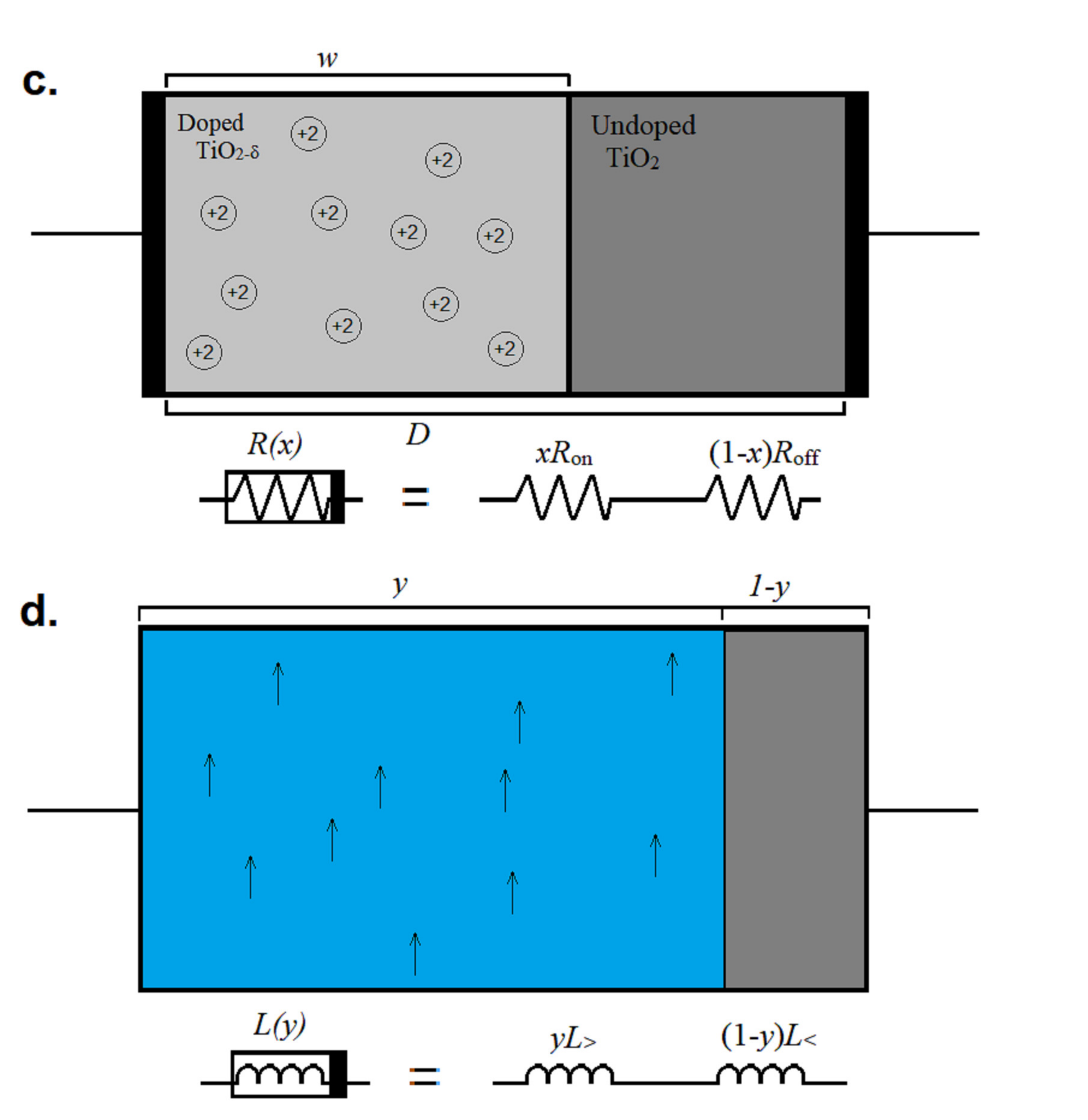}
\caption{Schematic of a $\mathcal{PT}$-symmetric dimer. (a) An $LC$ circuit with resistance $R$ (top) is coupled to another with negative resistance $-R$ (bottom) via a coupling inductor $L_c$. (b) The flow of four eigenvalues $\epsilon_\alpha$  of the Hamiltonian Eq.(\ref{eq:Heff}), that describes the coupled dimer as a function of gain-loss strength $\gamma$, is shown when $R(x)$ and $L_c(y)$ are constant. The first exceptional point occurs at $\gamma=\gamma_{PT}$ while the second one occurs at $\gamma=\gamma_c$. For a dimer with $\mu=1$, these values are $\gamma_\mathrm{PT}=0.732\omega_0$ and $\gamma_c=2.732\omega_0$, giving $\gamma_c=3.732\gamma_\mathrm{PT}$. (c) A simple model for memory resistor (memristor) is a doped, low resistance region in series with an undoped, high-resistance region, where the dynamics of the fractional width of the doped region $x(t)=w(t)/D$ are governed by dopant mobility, i.e. Eq.(\ref{eq:dopant_drift}). (d) A similar model is used for the meminductor, where the internal-state variable $y(t)$ denotes the linear fraction of the inductor core that is spin-polarized, Eq.(\ref{eq:y1}).}
\label{fig:schematic}
\end{figure*}

\section{Coupled $LC$ circuits with $\mathcal{PT}$ symmetry}
\label{sec:ptcircuits}
We start with the review of an electrical $\pt$-symmetric dimer~\cite{Schindler2011,Chitsazi2017,LeonMontiel2018,Wang2020}. Let us consider two identical $LC$ circuits, with effective, parallel resistors $\pm R$ respectively, that are connected with a coupling inductor $L_c$ as shown in Fig.~\ref{fig:schematic}a. When the two circuits are uncoupled, the energy in the standard $RLC$ circuit undergoes overdamped or underdamped decay, while the energy in the $-RLC$ circuit, with negative effective resistance, grows with time reflecting a time-reversed dynamics. The state of the coupled system is characterized by a real vector $\ket{\phi(t)}=[V_1(t),V_2(t),I_1(t),I_2(t),I_c(t)]^T$ where $V_{1(2)}(t)$ is the voltage across the first (second) capacitor, $I_{1(2)}(t)$ is the current through the first (second) inductor, and $I_c(t)$ is the current flowing through the coupling inductor $L_c$. Its equation of motion, determined by the Kirchhoff laws, can be written as $i\partial_t\ket{\phi(t)}=M\ket{\phi(t)}$ where the purely imaginary, $5\times5$ matrix $M$ of rank 4 is given by 
\begin{equation}
\label{eq:kirchoff}
M=i
\left[\begin{array}{ccccc}
-\frac{1}{RC} & 0  & -\frac{1}{C} & 0  & -\frac{1}{C}\\
 0 & +\frac{1}{RC} & 0  & -\frac{1}{C} & \frac{1}{C}\\
\frac{1}{L} & 0 & 0 & 0 & 0\\
0 & \frac{1}{L}  & 0 & 0 &  0\\
\frac{1}{L_c} & -\frac{1}{L_c} & 0 & 0 & 0
\end{array}\right].
\end{equation}

To map the Kirchhoff-law equations into the Schrodinger-like equation, we note that the circuit energy is given by $\mathcal{E}(t)=\bra{\phi(t)}A\ket{\phi(t)}$ where $A=\mathrm{diag}(C,C,L,L,L_c)/2$ is a diagonal matrix. Therefore, we consider the dynamics of energy density by defining $\ket{\psi}=A^{1/2}\ket{\phi}$ such that the norm of the state $\ket{\psi}$ is the total energy in the circuit. This mapping transforms the Kirchoff-law equations into 
\begin{equation}
\label{eq:se}
i\partial_t\ket{\psi}=H\ket{\psi},
\end{equation}
where the purely imaginary, rank-4 Hamiltonian $H_\mathrm{eff}=A^{1/2}MA^{-1/2}$ is given by
\begin{equation}
\label{eq:Heff}
H_\mathrm{eff}=i\omega_0
\left[\begin{array}{ccccc}
-\Gamma & 	0 		& 	-1 	& 	0 	& -\mu\\
0 		&	\Gamma 	& 	0 	& 	-1	& \mu\\
1		& 	0 		& 	0 	& 	0 	& 0\\
0 		& 	1 		& 	0 	& 	0 	& 0\\
\mu  	& 	-\mu	& 	0	& 	0 	& 0
\end{array}\right].
\end{equation}
Here, $\omega_0=1/\sqrt{LC}$ is the fundamental frequency of a single $LC$ circuit, $\mu=\sqrt{L/L_c}$ is the dimensionless inductive coupling between the lossy and active oscillators, and $\gamma=1/RC=\Gamma\omega_0$ is the gain-loss rate for the circuit. When $\gamma=0$, the Hamiltonian $H_\mathrm{eff}$ is Hermitian, and the norm of the state, i.e. the total energy in the circuit, is conserved. We also note that $H_\mathrm{eff}(\gamma)$ is $\pt$-symmetric with respect to  
\begin{equation}
\mathcal{P} = 
\left[\begin{array}{ccc}
	\sigma_{x} 	& 	0 			&	0\\
	0 			& 	\sigma_{x} 	&	0\\
	0 			& 	0 			& 	-1
\end{array}\right],\,
\mathcal{T} =U\mathcal{K}=
\left[\begin{array}{cc}
	\mathbbm{1}_{2} 	& 	0\\
	0 		& 	-\mathbbm{1}_{3}
\end{array}\right]\mathcal{K},
\label{eq:pt_ops}
\end{equation}
where $\sigma_{x}$ is the standard Pauli matrix and $\mathbbm{1}_{k}$ is a $k\times k$ identity matrix. $H_\mathrm{eff}$ also satisfies $\Pi H_\mathrm{eff}=-H_\mathrm{eff}\Pi$ where $\Pi=\mathcal{P}U$. This chiral symmetry is responsible for its spectrum consisting of a trivial zero and two pairs of particle-hole symmetric eigenvalues~\cite{Joglekar2010} given by
\begin{equation}
\label{eq:Heig}
\epsilon_\alpha=\pm\frac{\omega_0}{\sqrt{2}}\left[2+2\mu^2-\Gamma^2\pm\sqrt{(2\mu^2-\Gamma^2)^2-4\Gamma^2}\right]^{1/2}.
\end{equation}
It is easy to check that $\epsilon_\alpha$ are purely real for $\gamma\leq\gamma_\mathrm{PT}=\omega_0(\sqrt{1+2\mu^2}-1)$, and become purely imaginary when $\gamma>\gamma_c=\omega_0(\sqrt{1+2\mu^2}+1)$. Figure~\ref{fig:schematic}b shows the flow of the four eigenvalues in units of $\omega_0$ as a function of the gain-loss strength $\gamma/\gamma_\mathrm{PT}$ for a strongly coupled $LC$ circuit with $\mu=1$. 

When the dissipation $\gamma$ is time-dependent, the electrical-circuit dynamics still maps onto a Hamiltonian given by $H_\mathrm{eff}(t)=A^{1/2}M(t)A^{-1/2}$. On the other hand, if a conservative circuit element varies with time, the change-of-basis matrix $A^{1/2}(t)$ leads to a new Hamiltonian~\cite{LeonMontiel2018}
\begin{equation}
\label{eq:gauge}
H'_\mathrm{eff}=A^{1/2}(t)M(t)A^{-1/2}(t)+(i/2)\partial_t\ln A(t).
\end{equation}
Whether the additional gauge term in Eq.(\ref{eq:gauge}) commutes with the $\mathcal{PT}$ operator depends on the functional time-dependence of matrix $A(t)$. However, (with a little license with notation) we will continue to call such system a $\pt$-symmetric dimer. In the following sections, we will consider models where the time-dependence arises through memory in either the gain-loss strength $\gamma$ or the inductive coupling $\mu$. 

\section{ Memristive $\pt$-symmetric model}  
\label{sec:ptmemristor}

A memristor is a resistor whose resistance $R(x)$ depends on an internal, dimensionless state variable $x$~\cite{DiVentra2009}. The equation of motion for the state variable $x(t)$, in turn, is determined by the underlying microscopic model for the memristor. A memristor (memory resistor) was postulated by L. Chua more than half a century ago based on symmetry arguments~\cite{Chua1971}. It was realized just over a decade ago in a thin-film device with one monolayer of TiO$_2$ and its oxygen-vacancy doped counterpart, TiO$_{2-\delta}$~\cite{Strukov2008,Tour2008}. Since then it has become clear that memristive systems, with their  pinched-loop hysteresis signature~\cite{Chua2013} in the current-voltage characteristics, manifest themselves in semiconductor thin films, thermistors~\cite{Sapoff1963}, as well as ion channels~\cite{Hodgkin1952} in biological membranes~\cite{CHUA2012}. Here, we will consider the simplest model for its internal state variable~\cite{Strukov2008,Joglekar2009}. The TiO$_2$/TiO$_{2-\delta}$ thin film with size $D$ ($D\sim 5$ nm) can be modeled as two resistors in series, where the size of the doped part is given by $w=xD$, and the size of the undoped region is $D-w=(1-x)D$ (Fig.~\ref{fig:schematic}c). The resistance of this two-terminal passive device is given by 
\begin{equation}
\label{eq:mem1}
R(x)=xR_\mathrm{on}+(1-x)R_\mathrm{off},
\end{equation}
where $R_\mathrm{on}$ ($\sim$1 $\mathrm{k}\Omega$) is the resistance of the device if it is entirely doped and $R_\mathrm{off}\sim 10^2R_\mathrm{on}\gg R_\mathrm{on}$ is the resistance of the insulating TiO$_2$ device. When voltage is applied to such a device, in addition to conduction electrons, the charge +2 oxygen vacancies also move. The effect of their motion is amplified due to the two-monolayer thickness of TiO$_2$/TiO$_{2-\delta}$, and it determines the fractional width $x(t)$ of the doped region. By equating the rate of change of $x(t)$ with the drift velocity of the oxygen vacancy dopants, we get 
\begin{equation}
dx/dt=\eta(I/Q_0)F(x).
\label{eq:dopant_drift}
\end{equation}
Here $I$ is the electronic current through the memristor, $Q_0=D^2/\mu_DR_\mathrm{on}$ is the characteristic charge-scale for the memristor, the dopant mobility $\mu_{D}$ ($\sim10^{-10}$ cm$^{2}$V$^{-1}$s$^{-1}$) is 10-12 orders of magnitude smaller than the corresponding electron mobilities, and $\eta=\pm 1$ signifies whether the  doped region is shrinking or growing, i.e. polarity of the memristor.

The window function $F(x)$ in Eq.(\ref{eq:dopant_drift}) suppresses dopant mobility when the interface between undoped and doped regions approaches the device boundaries, i.e. $x\rightarrow 0$ or $x\rightarrow 1$. We use a family of window functions $F_p(x)=1-(2x-1)^{2p}$~\cite{Joglekar2009}. Since $x=\{0,1\}$ are fixed points of Eq.(\ref{eq:dopant_drift}), if the time-evolved state variable $x(t)$ reaches either fixed point, it has no further dynamics. However, the amount of charge required to change the state-variable value from $x=\delta_1\ll1$ to $x=\delta_2\ll\delta_1$ diverges as $(D^2/4p\mu_DR_\mathrm{on})\ln(\delta_1/\delta_2)$. Thus, starting from an $x\in(0,1)$ it is impossible to reach the boundaries in finite amount of time~\cite{Joglekar2012m}. In calculations, if $x(t_0+\delta t)$ reaches or exceeds the fixed-point values while $x(t_0)\in(0,1)$, either a smaller time-step $\Delta t$ is chosen or the updated $x$-value is shifted to just inside the boundaries to circumvent this numerical artifact. 

We now consider a $\pt$-symmetric dimer with a memristor $R(x)$ and a balanced gain-resistor that matches the instantaneous dissipation in the memristor. Such a setup is experimentally feasible with lock-in amplifiers and synthetic elements. The circuit dynamics now are described by two coupled, nonlinear, first-order equations, 
\begin{eqnarray}
\label{eq:heff1}
&i\partial_t\ket{\psi(t)}=H_\mathrm{eff}(\gamma(x))\ket{\psi(t)},\\
\label{eq:x1}
&\frac{dx}{dt}=\eta\frac{F(x)}{R(x)Q_0}V_1(t)=\eta\frac{F(x)}{R(x)Q_0}\sqrt{\frac{2}{C}}\langle 1|\psi(t)\rangle,
\end{eqnarray}
where $\gamma(x)=1/R(x)C$ denotes the variable gain-loss strength, and $\langle 1|\psi(t)\rangle$ is the first element of the energy-density state vector $|\psi(t)\rangle$.  Note that due to nonlinearities, the system dynamics depend on the initial state-norm or equivalently, the initial energy in the system. Therefore, the fate of the $\pt$-symmetric system with memory cannot be analyzed in terms of its state-variable dependent Hamiltonian $H_\mathrm{eff}(\gamma(x))$. Instead, we track the time-dependent energy $\mathcal{E}(t)=\langle\psi(t)|\psi(t)\rangle$ in the circuit. As in the standard $\mathcal{PT}$-symmetric case, this circuit energy $\mathcal{E}(t)$ shows bounded, periodic, oscillatory behavior for some parameter regime, while for others, it shows non-periodic, divergent behavior.  As we will demonstrate below, this behavior can sensitivity depend on the initial state $|\psi(0)\rangle$ as well. To characterize these two trends, we calculate the long-time amplification rate,
\begin{equation}
\label{eq:lyapunov}
\Lambda_\mathrm{amp}=\lim_{\tau\rightarrow\infty}\frac{1}{\tau}  \ln\left[\frac{\max\mathcal{E}(0\leq t\leq2\tau)}{\max\mathcal{E}(0\leq t\leq\tau)}\right].
\end{equation}
When the system is in the $\pt$-symmetric phase, the dynamics are oscillatory and therefore $\Lambda_\mathrm{amp}=0$. In the $\pt$-broken phase, due to the presence of amplifying modes, the energy grows exponentially with time and Eq.(\ref{eq:lyapunov}) characterizes its growth rate. Since the energy $\mathcal{E}(t)$ is obtained numerically, in practice, we must choose $\tau$ to be larger than any other relevant time-scale in the problem. 

For a $\pt$ symmetric circuit without memory, the fast timescales are given by $T_0=2\pi/\omega_0$ and $T_0/\mu^2\gg T_0$. The longest timescale is given by the inverse of the smallest eigenvalue difference, Eq.(\ref{eq:Heig}), and it diverges as one approaches the static $\pt$-threshold $\gamma_\mathrm{PT}$. Thus, one needs data at arbitrarily long time-scales to distinguish a system in the $\pt$-symmetric phase from one in the $\pt$-symmetry broken phase. For a memristor, the state-variable dynamics timescale depends on the dopant drift velocity or, equivalently, the applied voltage strength $v_0$, and is given by $T_\mathrm{m}=D^2/v_0\mu_D$~\cite{Joglekar2009}. Equating these two time scales, we obtain the characteristic voltage scale $v_0=D^2\omega_0/2\pi\mu_D$; this voltage gives rise to a linear drift of size $D$ in one oscillation of the $LC$ circuit. Note that since the memristor value is confined between $R_\mathrm{on}\leq R(x)\leq R_\mathrm{off}$, the dissipation strength $\gamma(x)$ is also bounded between $\gamma_\mathrm{off}$ and $\gamma_\mathrm{on}$. If both strengths are below $\gamma_\mathrm{PT}$, the time-averaged dissipation, defined as 
\begin{equation}
\label{eq:avggamma}
\bar{\gamma}(t)=\frac{1}{t}\int^{t}_0\gamma(x(t'))\, dt'
\end{equation}
will be smaller than the threshold {\it for any t}. Then, in the absence of any Floquet resonances~\cite{Joglekar2014,Lee2015}, the system dynamics will be bounded and oscillatory.  Similarly, if both are above $\gamma_\mathrm{PT}$, so is the time-averaged dissipation, and with the same caveat, the system will be in the $\pt$-symmetry broken region, as indicated by a divergent circuit energy growth. Therefore, for simulations, we choose $0<\gamma_\mathrm{off}\leq\gamma_\mathrm{PT}$ and $\gamma_\mathrm{PT}\leq\gamma_\mathrm{on}\leq\gamma_c$. 

\begin{figure}
\centering
\includegraphics[width=\columnwidth]{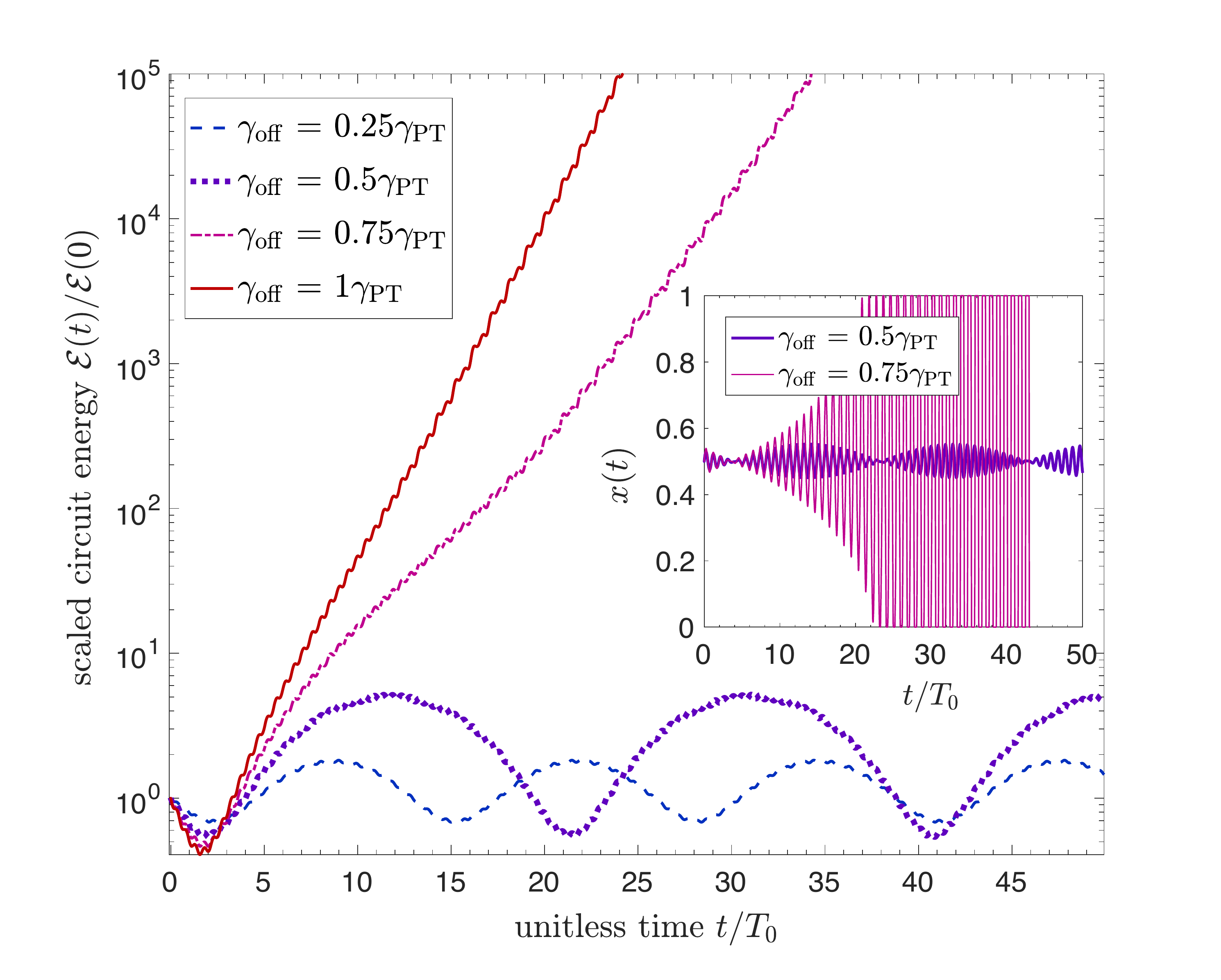}
\caption{Scaled circuit energy $\mathcal{E}(t)/\mathcal{E}(0)$ as a function of unitless time $t/T_0$ shows transition from oscillatory behavior at small dissipation $\gamma_\mathrm{off}/\gamma_\mathrm{PT}\leq 0.5$ to exponential growth at large dissipation, $\gamma_\mathrm{off}/\gamma_\mathrm{PT}=\{0.75,1\}$. The fast oscillations with period $T_0=2\pi/\omega$ on top of the slow dynamics are present in both phases. The circuit parameters are $\gamma_\mathrm{on}=2\gamma_\mathrm{PT}$, $\mu=0.3$, $x(0)=0.5$, and $|\psi(0)\rangle=|\psi_1\rangle$ with a small initial voltage $V_1(0)=0.5v_0$. Inset: the internal state variable shows small-amplitude oscillations about its initial value in the $\mathcal{PT}$-symmetric phase, whereas in the $\mathcal{PT}$-broken phase, $x(t)$ oscillates while reaching both of its extrema.}
\label{fig:mem1}
\vspace{-5mm}
\end{figure}

\begin{figure*}
\centering
\includegraphics[width=0.495\textwidth]{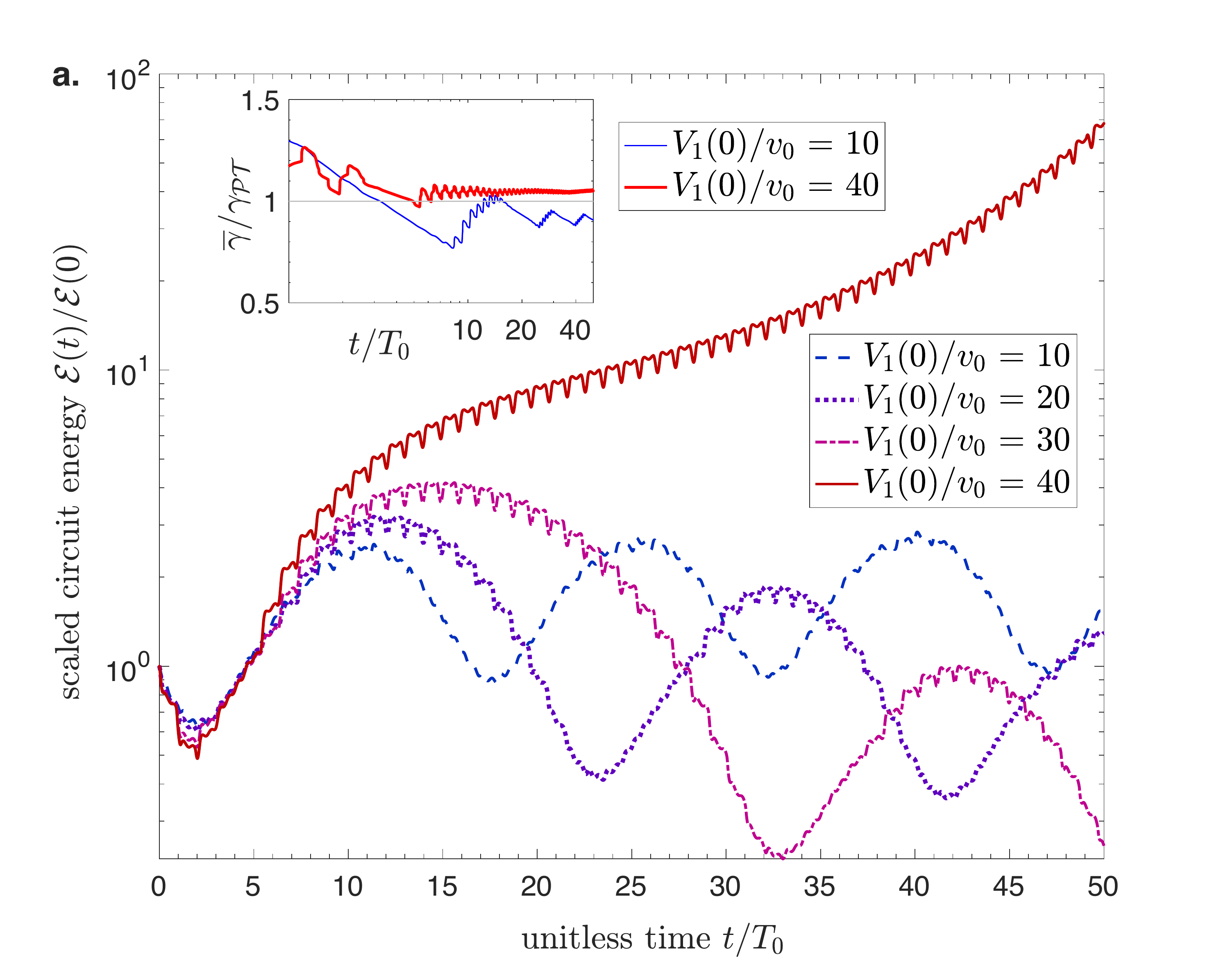}
\includegraphics[width=0.495\textwidth]{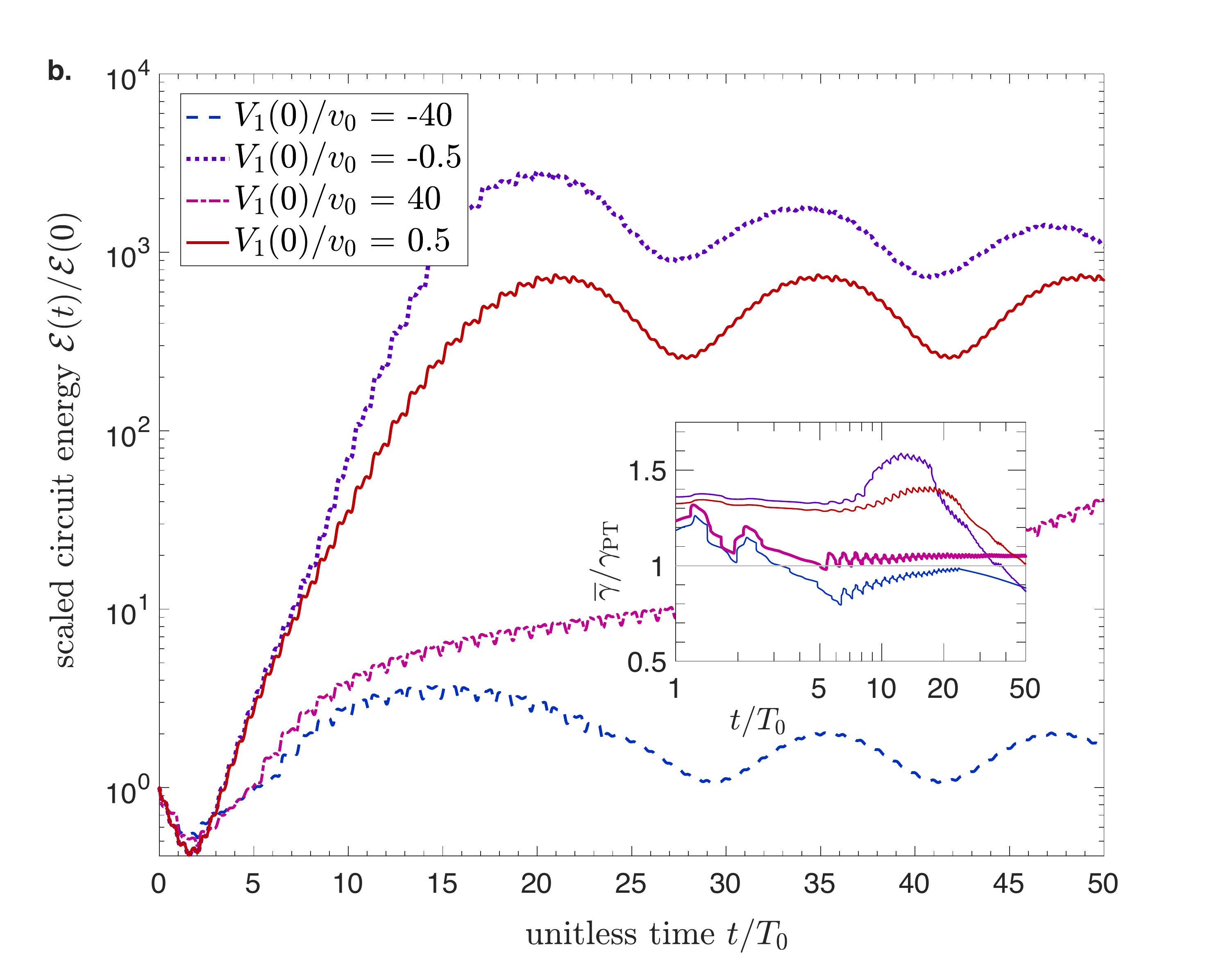}
\includegraphics[width=0.495\textwidth]{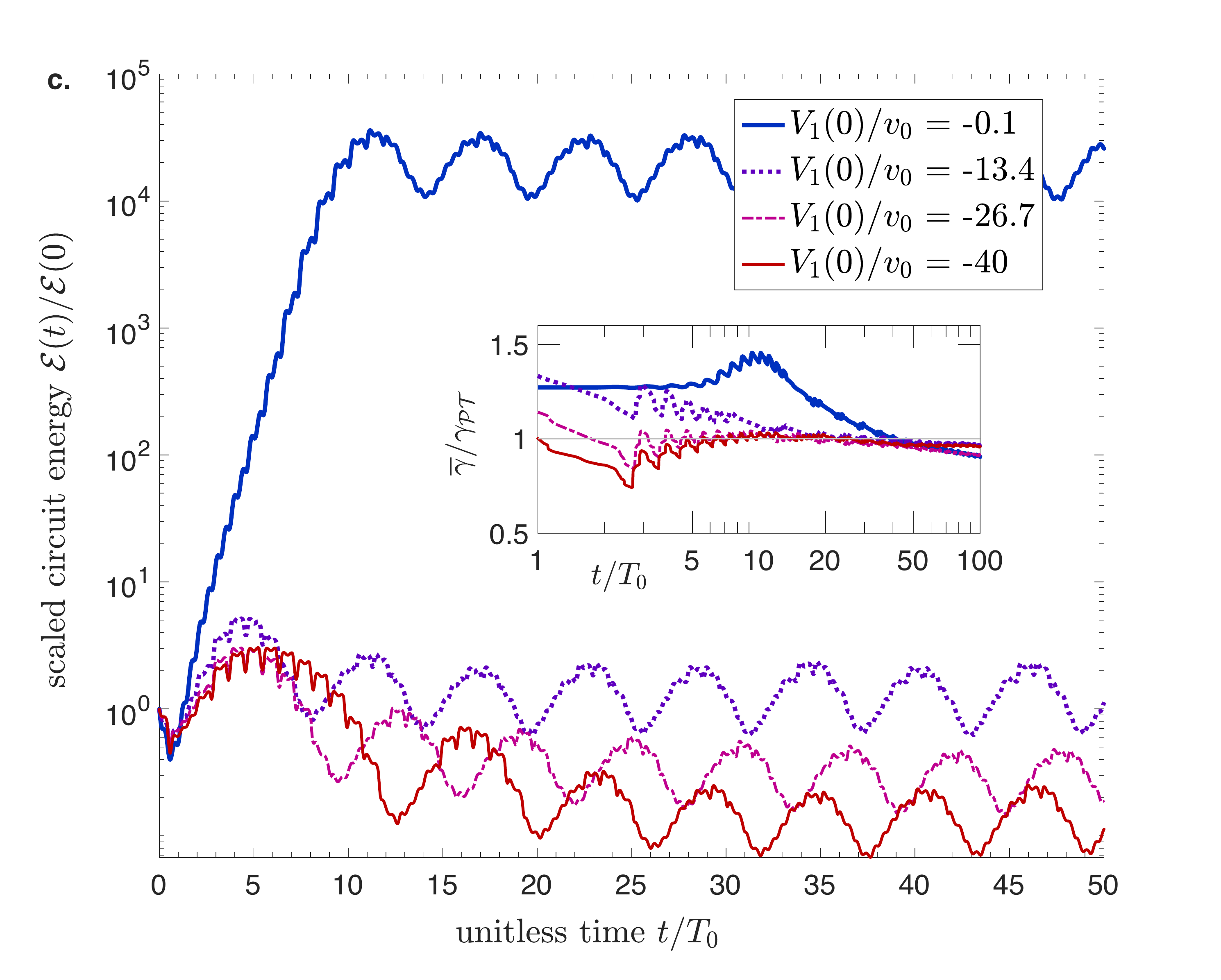}
\includegraphics[width=0.495\textwidth]{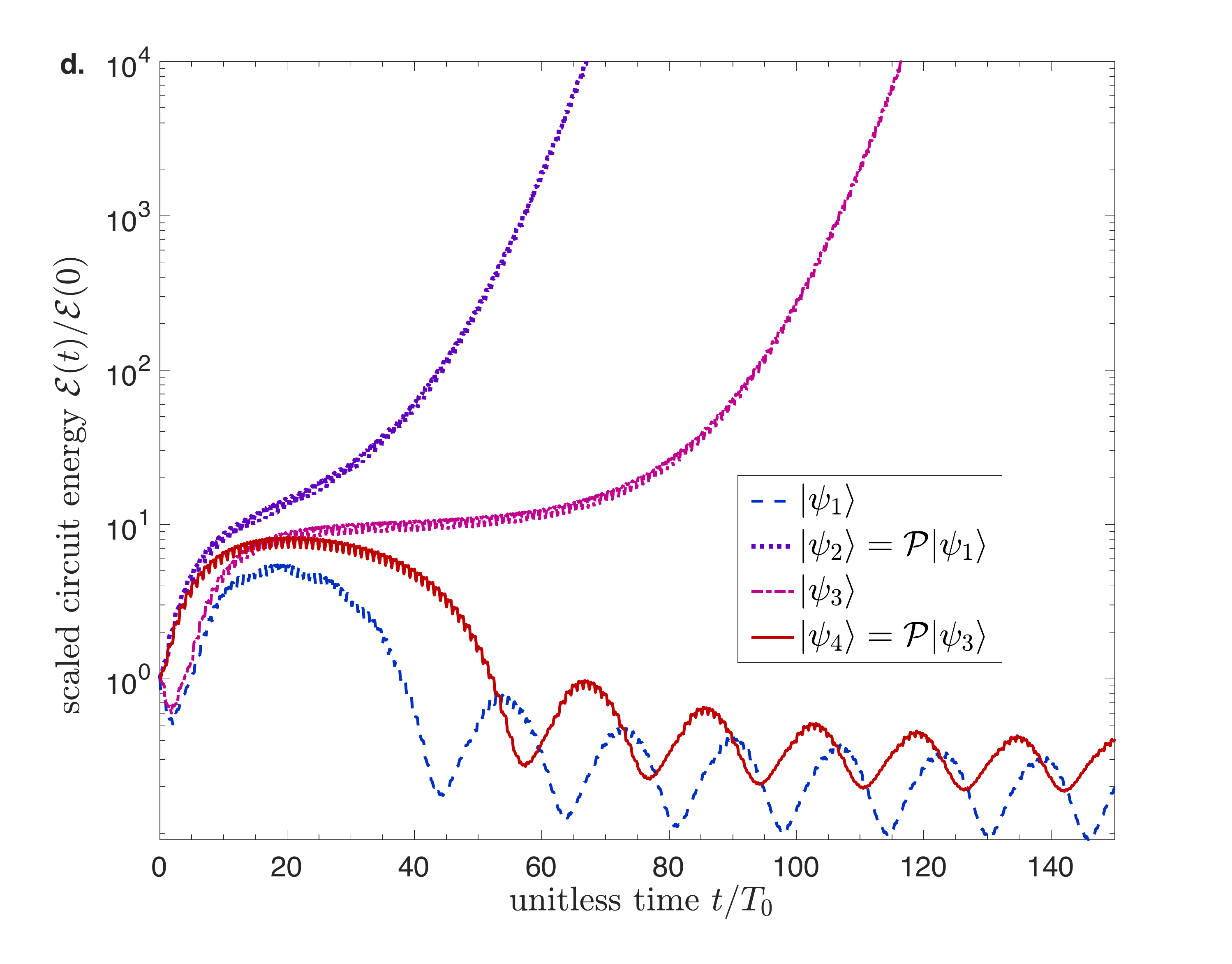}
\caption{Non-trivial dynamics of a $\pt$-symmetric dimer with memristive dissipation. (a) Depending on the initial state norm $\mathcal{E}(0)$, the scaled circuit energy shows oscillations or exponential growth. (b) At small initial voltages $V_1(0)=\pm 0.5v_0$, the circuit energy shows oscillatory behavior after an exponential growth transient. At large initial voltages, $V_1(0)=\pm 40v_0$, the scaled circuit energy either oscillates without notable amplification ($|\psi(0)\rangle=+80|\psi_1\rangle$) or grows exponentially ($|\psi(0)\rangle=-80|\psi_1\rangle$). (c) For the same initial energy, changing the initial state from $\alpha|\psi_1\rangle$ to $-\alpha|\psi_1\rangle$ shows qualitatively different energy dynamics; here $0.2\leq\alpha\leq 80$ spans all different initial states considered. (d) Initial states $75|\psi_1\rangle$ and $75\mathcal{P}|\psi_1\rangle$ with same norm show oscillatory and exponential behavior, as do states $|\psi_4\rangle$ and $|\psi_3\rangle=\mathcal{P}|\psi_4\rangle$, respectively. Insets: the average dissipation strength $\overline{\gamma}(t)/\gamma_\mathrm{PT}$ is less than unity at long times for oscillatory behavior, and exceeds unity at long times for exponential behavior. Rest of the system parameters for results in (a)-(d) are given in the text.}
\label{fig:mem2}
\end{figure*}

Figure~\ref{fig:mem1} shows the typical energy dynamics $\mathcal{E}(t)$ for a memristive dimer as a function of $\gamma_\mathrm{off}$. The circuit parameters are $\mu=0.3$, $\gamma_\mathrm{on}=2\gamma_\mathrm{PT}$, and $x(0)=0.5$. The initial state vector $|\psi(0)\rangle$ of the system is given by 
\begin{eqnarray}
\label{eq:psi1}
|\psi_1\rangle &=&\sqrt{C/2}[V_1(0),0,0,0,0]^T,\\
V_1(0) &=&0.5v_0.
\end{eqnarray}
At low dissipation strength, $\gamma_\mathrm{off}=\{0.25,0.5\}\gamma_\mathrm{PT}$, the scaled circuit energy shows bounded oscillations whose period increases as the dissipation is increased. At large dissipations,  $\gamma_\mathrm{off}=\{0.75,1\}\gamma_\mathrm{PT}$, the scaled circuit energy shows exponential growth that is characteristic of a $\mathcal{PT}$-symmetry broken phase. The inset shows the corresponding dynamics of the internal state variable $x(t)$. In the $\mathcal{PT}$-symmetric phase, the doped fraction $x(t)$ undergoes small oscillations around its mean value, whereas in the $\mathcal{PT}$ broken phase, it reaches its extremal values, thus driving the dissipation between $\gamma_\mathrm{on}$ and $\gamma_\mathrm{off}$ over the timescale $T_0$. The results in Fig.~\ref{fig:mem1} are, for the most part, expected even for a $\mathcal{PT}$-symmetric dimer {\it without memory.} In the following paragraphs, we will show the unique features that arise from the memristive nature of dissipation and the resultant non-linearity in this system. 

Figure~\ref{fig:mem2}a shows the temporal evolution of circuit energy for four initial states $|\psi(0)\rangle=\{20,40,60,80\}|\psi_1\rangle$, with circuit parameters $\mu=0.3$, $\gamma_\mathrm{on}=2\gamma_\mathrm{PT}$, $\gamma_\mathrm{off}=0.3\gamma_\mathrm{PT}$, and $x(0)=0.5$. At small initial energies $\mathcal{E}(0)=\langle\psi(0)|\psi(0)\rangle$, the dynamics are oscillatory with a period that increases with increasing energy; this behavior changes over to exponential growth when $|\psi(0)\rangle=80|\psi_1\rangle$. In a linear $\mathcal{PT}$-symmetric dimer -- static or Floquet -- all of these states are equivalent to $|\psi_1\rangle$ and will exhibit identical results for the scaled circuit energy $\mathcal{E}(t)/\mathcal{E}(0)$. Instead, due to the nonlinearity introduced by dynamics of the internal state variable, now the fate of the system depends on the initial circuit energy.

Figure~\ref{fig:mem2}b shows the diverse energy dynamics that occur for four initial states given by $|\psi(0)\rangle=\pm |\psi_1\rangle, \pm 80|\psi_1\rangle$; the initial doped fraction is $x(0)=0.9$ and rest of the circuit parameters are the same as in Fig.~\ref{fig:mem2}a. We see that the fate of the scaled circuit energy depends not only on the initial energy, but also on the sign of the initial voltage $V_1(0)$, or equivalently, the polarity $\eta=\pm1$ of the memristor. This can also be interpreted as the phase of the initial state $|\psi(0)\rangle$. (Recall that the phase of the purely real state-vector $|\psi(t)\rangle$ is restricted to 0 or $\pi$). Specifically, at small initial energies, i.e. $V_1(0)=\pm 0.5v_0$, the scaled energy shows an exponential-growth transient followed by an oscillatory behavior that persists at long times $t/T_0\sim 10^3$ (not shown).  At higher initial energy, the system starting in the initial state $|\psi(0)\rangle=-80|\psi_1\rangle$ shows a clear oscillatory energy dynamics with minimal net amplification, whereas with initial state $+80|\psi_1\rangle$, the scaled circuit energy undergoes exponential amplification. 

Figure~\ref{fig:mem2}c shows results for a moderately coupled dimer with $\mu=0.5$, $\gamma_\mathrm{on}=2\gamma_\mathrm{PT}$, $\gamma_\mathrm{off}=0.3\gamma_\mathrm{PT}$, $x(0)=0.9$, and four initial states given by $|\psi(0)\rangle=\{-0.2,-26.8,-53.4,-80\}|\psi_1\rangle$. As the negative prefactor of $|\psi_1\rangle$ is increased in magnitude, we see that the scaled circuit energy $\mathcal{E}(t)/\mathcal{E}(0)$ settles into oscillations about a mean that is monotonically suppressed. When the initial circuit energy $\mathcal{E}(0)$ is small, $V_1(0)/v_0=-0.1$, the scaled energy starts oscillating after an exponential growth transient similar to that seen in panel b for small initial voltages. It is important to note that this behavior also contrasts with Fig.~\ref{fig:mem2}a, where the largest $V_1(0)$ value resulted in a $\mathcal{PT}$ symmetry broken phase. 

Lastly, in Fig.~\ref{fig:mem2}d, we explore the behavior of the circuit energy for four initial states given by $75|\psi_1\rangle$, $|\psi_2\rangle=75\mathcal{P}|\psi_1\rangle$, $|\psi_3\rangle=\sqrt{L/2}[0,0,I_1(0),0,0]^T$, and $|\psi_4\rangle=\mathcal{P}|\psi_4\rangle$. The initial voltages $V_1(0)=V_2(0)=35v_0$ or initial currents $I_1(0)=I_2(0)$ are chosen such that all states have the same initial energy (or state norm); the rest of the circuit parameters are the same as those for Fig.~\ref{fig:mem2}a. We see that when the system starts with nonzero voltage on the dissipative $LC$ circuit ($75|\psi_1\rangle$), after a growth transient, the scaled circuit energy oscillates with a mean that is below the initial circuit energy. In contrast, when the system starts with nonzero voltage on the amplifying $LC$ circuit ($|\psi_2\rangle=75\mathcal{P}|\psi_1\rangle$), the scaled energy diverges exponentially indicating a $\pt$-symmetry broken state. In contrast, when we start with a nonzero inductor current in the dissipative $LC$ circuit ($\psi_3\rangle$), the scaled energy shows exponential growth, while switching the nonzero initial current to the amplifying $LC$ circuit ($|\psi_4\rangle=\mathcal{P}|\psi_3\rangle$) leads to stable oscillatory dynamics for the scaled circuit energy. These properties are dramatically different from those of a traditional, memory-less $\pt$-symmetric system. In the latter, different distributions of the initial energy density only introduce temporal shift in the dynamics of scaled circuit energy. 

The insets in Fig.~\ref{fig:mem2} show the evolution of the average dissipation strength $\overline{\gamma}(t)$ relative to the $\pt$-breaking threshold strength $\gamma_\mathrm{PT}$ in a $\pt$-symmetric dimer with no memory. A common pattern observed is that when their ratio is less than unity at long times, the scaled circuit energy shows oscillatory behavior, whereas if the ratio exceeds unity at long times, the scaled circuit energy grows exponentially. In the following section, we will show that this naive expectation is wrong. 


\section{$\pt$ system with self-organized Floquet dynamics}
\label{sec:floquet}

In realistic memristors, the resistance of the undoped region can be orders of magnitude higher than that of the doped regions. When the initial state voltage is sufficiently high, the internal state variable $x(t)$ switches periodically between 0 and 1 at frequency $\omega_0$ as shown in the inset in Fig.~\ref{fig:mem1}. It implies that the gain-loss strength in the $\pt$-symmetric dimer effectively switches on and off, with $\gamma_\mathrm{off}\ll\gamma_\mathrm{on}$, i.e. the system behaves as if it has a time-periodic gain-loss strength whose periodicity is given by $T_0=2\pi/\omega_0$. 

It is known that in {\it memory-less} $\pt$-symmetric systems with periodic non-Hermiticity, the landscape of exceptional points that separate the $\pt$-symmetric phase from the $\pt$-symmetry broken phase is dramatically altered relative to its static limit~\cite{Joglekar2014,Lee2015}. In particular, the $\pt$-broken phase occurs at vanishingly small gain-loss strengths when the modulation frequency $\omega_0$ is an odd sub-harmonic of the Hermitian energy gap $\Delta\equiv\omega_0(\sqrt{1+2\mu^2}-1)$ of the system~\cite{Li2019a,Chitsazi2017,LeonMontiel2018}, 
\begin{equation}
\label{eq:floquetgamma}
\frac{\omega_0}{\Delta}=\frac{1}{(\sqrt{1+2\mu_n^2}-1)}=\frac{1}{(2n+1)},
\end{equation} 
where $n\geq0$. This analogy suggests that at $\mu_n=\sqrt{(2n+3)(2n+1)}/2$, the memristive $\pt$-symmetric system might show similar properties. The strongest $\pt$-symmetry broken region should then occur at the first resonance $\mu_0=\sqrt{3/2}\approx1.225$ for a vanishingly small gain-loss strength $\gamma_\mathrm{off}\ll\gamma_\mathrm{on}\lesssim\gamma_\mathrm{PT}$. 

\begin{figure*}
\centering
\hspace{-5mm}
\includegraphics[width=0.35\textwidth]{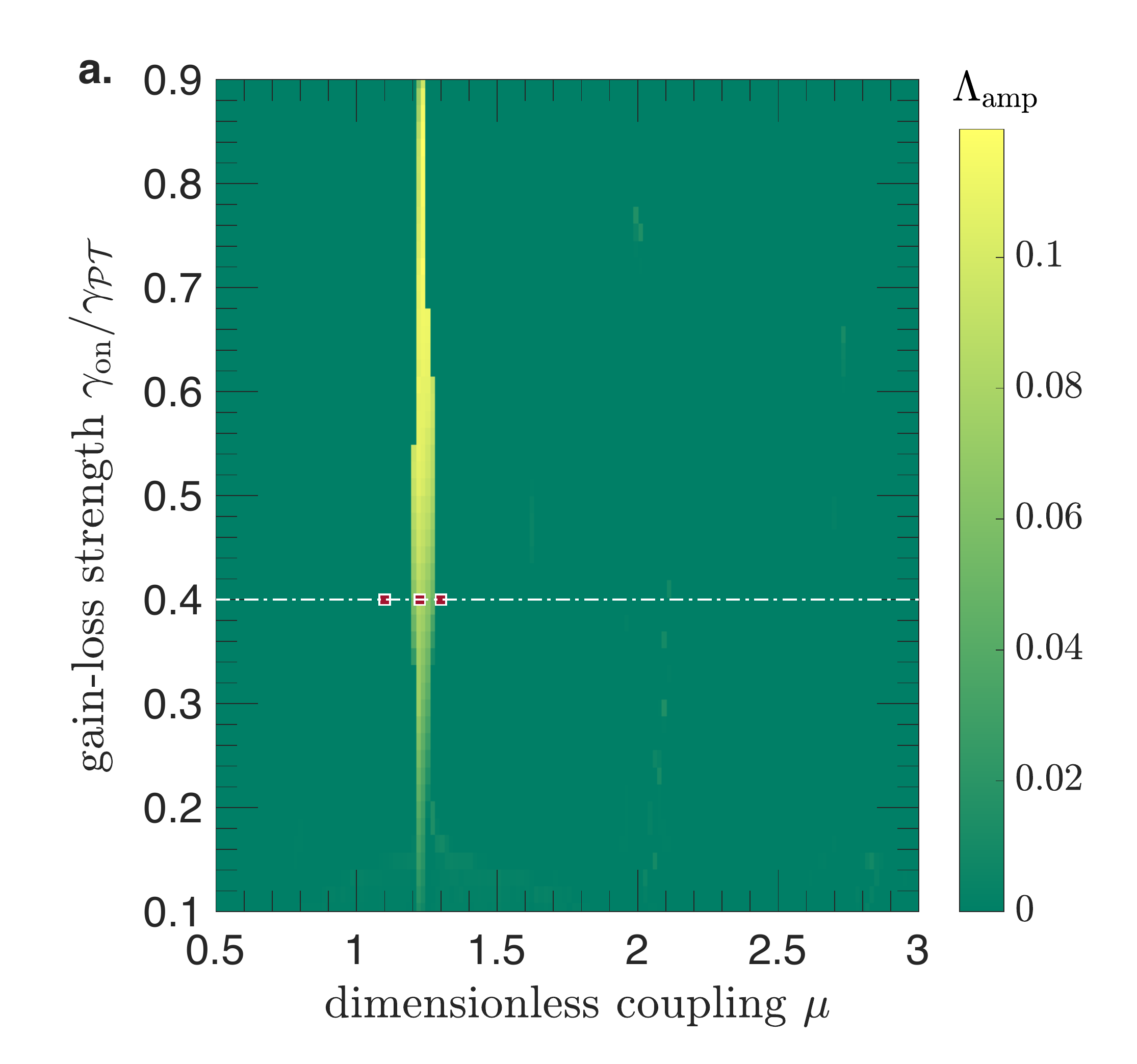}\hspace{-5mm}
\includegraphics[width=0.35\textwidth]{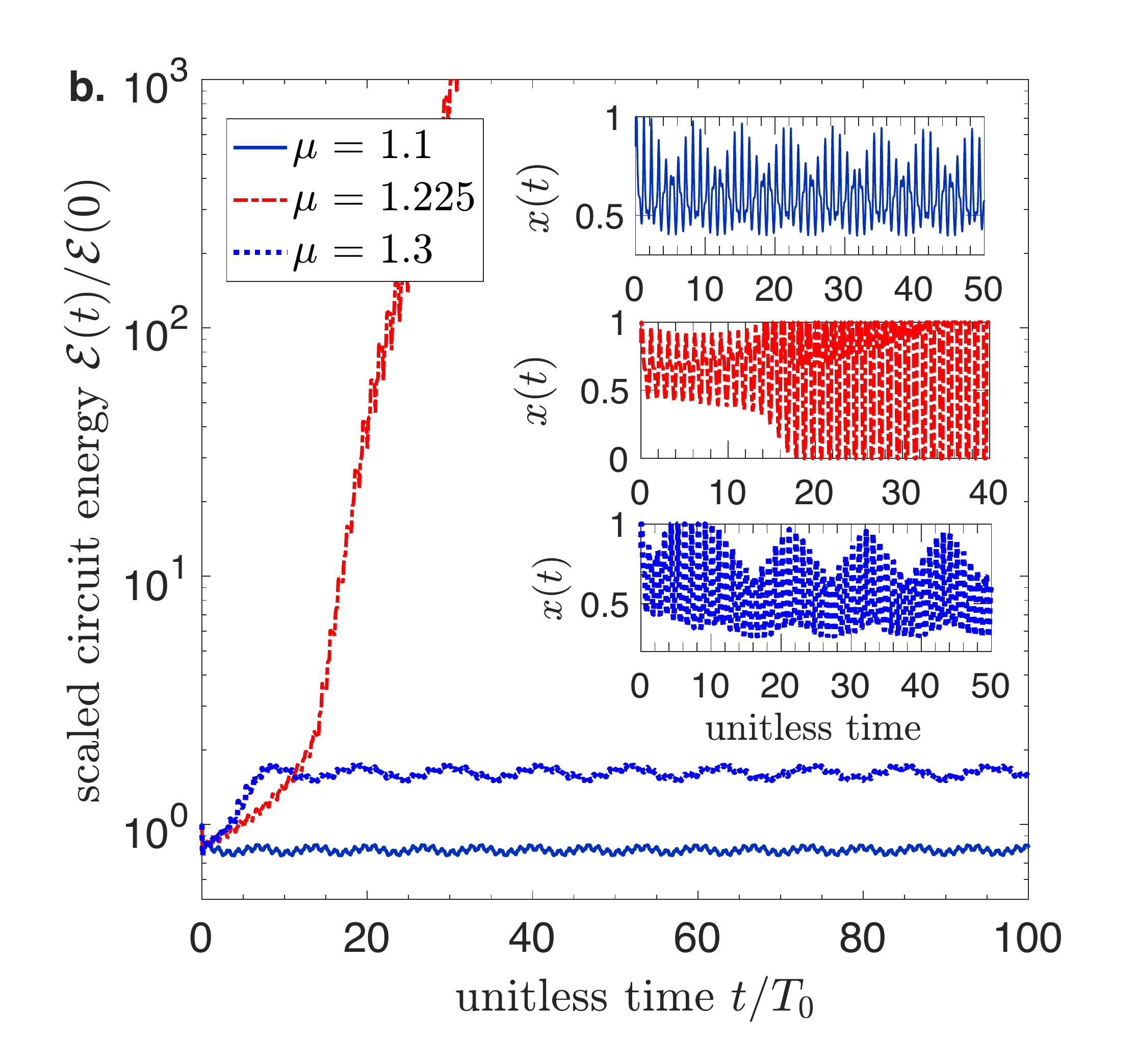}\hspace{-5mm}
\includegraphics[width=0.35\textwidth]{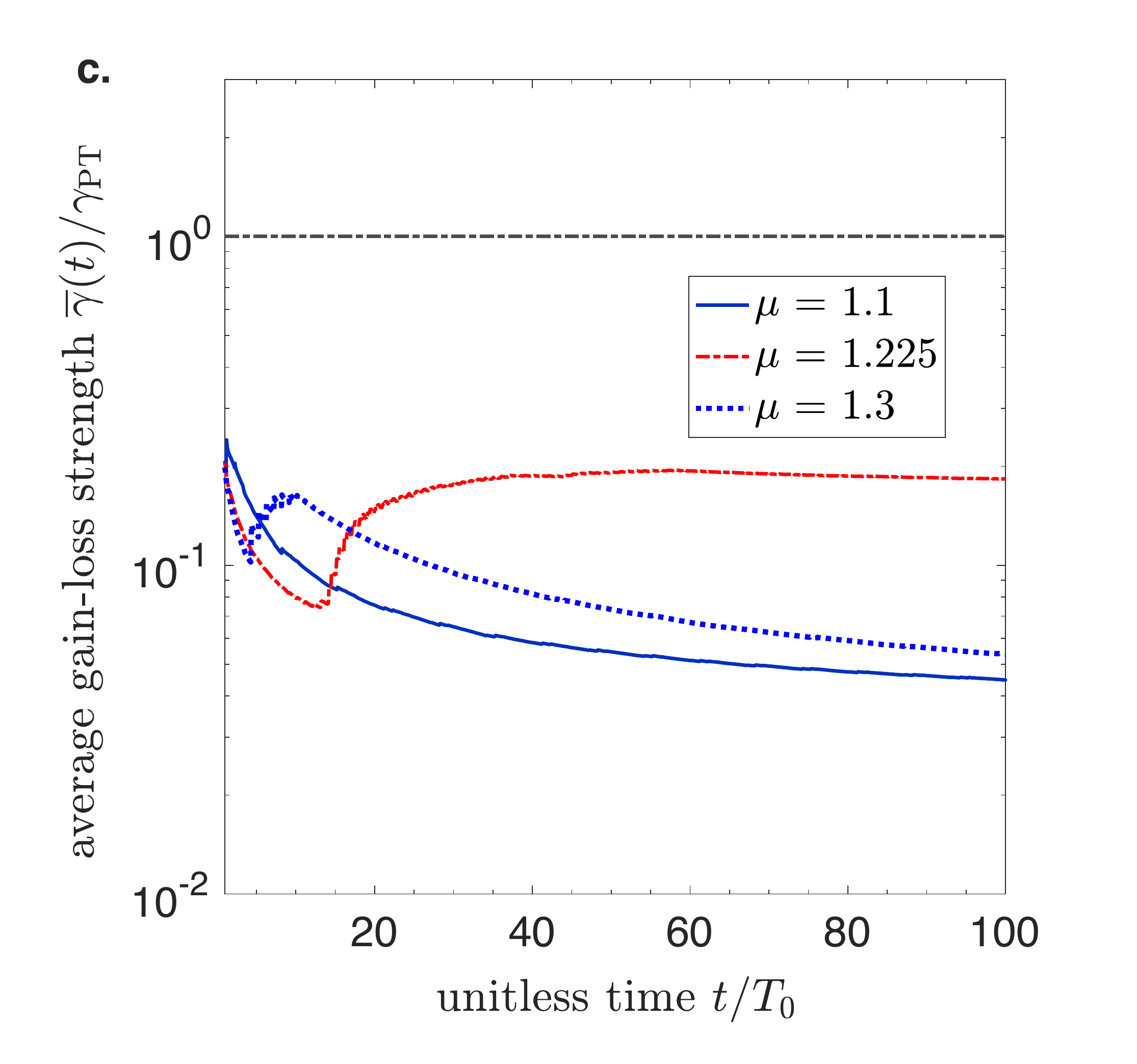}
\caption{Self-induced Floquet dynamics. (a) Plot of $\Lambda_\mathrm{amp}$, Eq.(\ref{eq:lyapunov}), in the small gain-loss and large coupling ($\mu\geq 1$) domain shows a vertical, $\pt$-symmetry broken region near $\mu_0\approx1.225$. Three red squares along the white dot-dashed line ($\gamma_\mathrm{on}/\gamma_\mathrm{PT}=0.4$) are at $\mu=\{1.1,\sqrt{3/2},1.3\}$. (b) The scaled circuit energy shows oscillatory behavior at $\mu=1.1$; it changes over to exponential at $\mu=1.225$, and back to amplified, oscillatory behavior at $\mu=1.3$. Inset: the internal state variable $x(t)$ oscillates between both extrema in the $\pt$-broken phase; in the $\pt$-symmetric phase, $x(t)$ oscillations do not reach small values. (c) Time-averaged gain-loss strength $\overline{\gamma}(t)$ at each of the three $\mu$ values remains well below the static $\pt$-threshold (black dot-dashed line), showing a key feature of $\pt$-symmetry breaking due to Floquet modulation of gain-loss strength.}
\label{fig:memfloquet}
\end{figure*}

We therefore carry out simulations with Eqs.(\ref{eq:heff1})-(\ref{eq:x1}) for circuit parameters $\gamma_\mathrm{off}=0.01\gamma_\mathrm{PT}$, $0.1\leq\gamma_\mathrm{on}/\gamma_\mathrm{PT}\leq 0.9$; initial doped fraction $0.1\leq x_0\leq 0.9$; and different initial states and memristor polarities. These initial conditions ensure that time-averaged dissipation remains below the static $\pt$-symmetry breaking threshold at all times, i.e. $\overline{\gamma}(t)<\gamma_\mathrm{PT}$. 

Figure~\ref{fig:memfloquet}a shows the typical plot for the amplification factor $\Lambda_\mathrm{amp}$, Eq.(\ref{eq:lyapunov}), in the $\mu-\gamma_\mathrm{on}$ plane. It is obtained with $|\psi(0)\rangle=70|\psi_1\rangle$, $x_0=0.85$, and $\tau=100T_0$. We see that the system is in the $\pt$-symmetric state ($\Lambda_\mathrm{amp}=0$) for most of the region except in the vicinity of $\mu_0=\sqrt{3/2}=1.225$, where the amplification factor is positive and grows with increasing $\gamma_\mathrm{on}$. These qualitative features remain the same for different initial circuit energies, state variable values $x_0$, and memristor polarities. Figure~\ref{fig:memfloquet}b shows the temporal evolution of the scaled circuit energy for $|\psi(0)\rangle=70|\psi_1\rangle$, $\gamma_\mathrm{on}=0.4\gamma_\mathrm{PT}$, and three different couplings marked by squares in Fig.~\ref{fig:memfloquet}a. At $\mu=1.1$, the system is in the $\pt$-symmetric phase with an approximately constant total energy. At slightly higher coupling $\mu\approx\mu_0$, the scaled energy shows a clear exponential growth indicating a $\pt$-symmetric broken regime. As the coupling is increased further to $\mu=1.3$, the system again enters the $\pt$-symmetric phase, albeit with an enhanced mean for the scaled circuit energy. The inset in Fig.~\ref{fig:memfloquet}b shows the dynamics of the internal state variable $x(t)$ at the three coupling values. In the $\pt$-symmetric region ($\mu=1.1$), the doped fraction oscillates without reaching both extrema; at $\mu=1.225$, this changes to square-wave oscillations between the two extema; and at $\mu=1.3$, the system is back in the $\pt$-symmetric phase and the $x(t)$ oscillation range is reduced again. 

Figure~\ref{fig:memfloquet}c shows corresponding time-averaged gain-loss strengths $\overline{\gamma}(t)$. We see that at $\mu=1.1$ and $\mu=1.3$, the average $\gamma$ saturates to $\sim5\%$ of the static threshold value, whereas for $\mu=1.225$, it saturates towards $\sim20\%$ of the static threshold. In all cases, however, it is significantly smaller than unity. This emergence of a positive amplification factor at specific couplings and very small gain-loss strengths is a key hallmark of Floquet $\pt$-symmetry breaking phenomenon~\cite{Li2019a,Joglekar2014,Lee2015,Chitsazi2017,LeonMontiel2018}. It can be qualitatively understood as follows: the energy oscillates between the lossy $LC$ unit and the active $LC$ circuit at frequency $\Delta(\mu)$ that is governed by the dimensionless coupling. If it is on the gain unit when $x=1$ and moves to the loss unit when $x=0$, the circuit energy will continue to amplify even if the gain-loss strengths $\gamma_\mathrm{off}$ and $\gamma_\mathrm{on}$ are small relative to $\gamma_\mathrm{PT}$. However, this synchronization requires the  memristor switching frequency to match the energy sloshing frequency, i.e. $\mu=\sqrt{3/2}$. Our results for the $\pt$-broken phase at small time-averaged $\gamma$ in the vicinity of the primary resonance remain qualitatively unchanged irrespective of initial circuit parameters, the initial state energy, the memristor polarity, or a different distribution of energy density across different elements in the initial state. As is expected, the nonzero amplification factor increases with $\gamma_\mathrm{on}$, increases with $\mathcal{E}(0)$, and decreases with initial fraction of the doped region $x(0)$. We also note that numerically, we do not find any evidence for $\pt$-broken region at higher values of $\mu$, suggesting the absence of such effect at odd sub-harmonics. 

The wide range of dynamics displayed by the circuit energy in a memristive $\pt$-symmetric dimer raises several considerations. First, the approximation of a constant negative resistor $-R$ or memristor $-R(x)$ breaks down at sufficiently large net amplification $\mathcal{E}(t)/\mathcal{E}(0)$, just as the approximation of a constant gain coefficient breaks down in the optical domain. Therefore, in reality, the exponential growth in the circuit energy will saturate and reach a steady-state value in the $\pt$-symmetry broken region, just as it does in the optical domain. Second, due to the very large parameter space in Eqs.(\ref{eq:heff1})-(\ref{eq:x1}), gaining a global understanding of whether the trajectories of $|\psi(t)\rangle$ in the four-dimensional space $\mathbbm{R}^4$ exhibit closed orbits, fixed points, open diverging orbits, or chaotic behavior is difficult. On the other hand, with increase in the studies of $\pt$-symmetric electrical lattice models~\cite{Hoffmann2020} and the easy availability of memristor emulators~\cite{HKim2012}, experimental investigation of these systems seems highly feasible. 

\section{$\pt$-symmetric dimer with meminductive coupling}
\label{sec:ptmeminductor}

A meminductor (memory inductor) is a two-terminal passive device whose inductance $L_c(y)$ depends on a dimensionless state variable $y$ whose dynamics, in turn, are governed by the current $I_c$ flowing through the inductor~\cite{Ventra2009}. Such a device shows a pinched hysteresis loop in the plane spanned by the current $I_c$ and the time-integral of the voltage (called the flux) $\phi$ across the device~\cite{ZYin2015}. The simplest, intuitive model of a meminductor is a solenoid with a ferromagnetic rod that can move in and out of its core~\cite{Slade2005}. However, the motion of the rod depends on the current $I_c$ through a second-order derivative, i.e. $d^2y/dt^2\propto I_c$~\cite{Slade2005}, and sets it apart from the viscous, drift-velocity model for the internal state variable of a memristor, Eq.(\ref{eq:dopant_drift}). 

To circumvent this distinction, we only consider meminductors in which  the internal state variable $y(t)$ has a viscous, drift dynamics~\cite{Ventra2009}, i.e. 
\begin{eqnarray}
\label{eq:memi1}
\phi& =& L_c(y) I_c,\\
dy/dt & = & f(y) I_c,
\end{eqnarray}
where the function $f(y)$ depends on the physical realization of a meminductor and corresponding system parameters. Inspired by the ferromagnetic-rod example, we consider $L_c(y)=yL_>+(1-y)L_<$ where $y(t)$ is the fractional size of the ``effective magnetic core'' (Fig.~\ref{fig:schematic}d). One microscopic mechanism for generating a current-induced spin polarization (or magnetization) is the spin Hall magnetoresistance effect~\cite{Dyakonov2007,Miao2014}, that enabled the realization of the meminductor in a platinum yttrium-iron-garnet (Pt/YIG) hybrid structure~\cite{Han2014}. We note that since Eq.(\ref{eq:memi1}) relates the flux to the current, the Kirchhoff-law equation for the current through the inductor is modified to      
\begin{equation}
\label{eq:newic}
\frac{dI_c}{dt}=\frac{(V_1-V_2)}{L_c}-\frac{\Delta L}{L_c}\frac{dy}{dt}I_c,
\end{equation}
where $\Delta L=L_>-L_<$ is the maximum change in the inductance. It is worth its while to emphasize that the second, $I_c$-dependent term in Eq.(\ref{eq:newic}) is absent for an inductor without memory. Its presence changes the sign of the gauge term introduced by a time-dependent change of basis, Eq.(\ref{eq:gauge}), giving rise to a new, effective Hamiltonian $\overline{H}_\mathrm{eff}=H_\mathrm{eff}(\mu(y))-(i/2)\partial_t\ln A(y)$. Thus, to investigate energy dynamics in a meminductive dimer, we solve the following set of coupled, nonlinear equations,
\begin{eqnarray}
\label{eq:heff2}
&i\partial_t\ket{\psi(t)}=\overline{H}_\mathrm{eff}(\mu(y))\ket{\psi(t)},\\
\label{eq:y1}
&\frac{dy}{dt}=\eta\frac{F(y)}{Q_c}I_c(t)=\eta\frac{F(y)}{Q_c}\sqrt{\frac{2}{L_c(y)}}\langle 5|\psi(t)\rangle,
\end{eqnarray}
where $\eta$ is the polarity of the meminductor, $F(y)$ is a window function that suppresses the change in $y(t)$ near its fixed points, and $Q_c$ is the material-dependent characteristic charge that generates sufficient spin accumulation to change $y$ from zero to unity. The corresponding characteristic current-scale for the coupling meminductor is given by $i_{0}=\omega_0Q_c$. This phenomenological drift model, Eq.(\ref{eq:y1}), produces key meminductor features such as a pinched loop hysteresis in the $\phi-I_c$ plot for an alternating current input~\cite{Ventra2009,Han2014}. 

For a pair of $\pm RLC$ circuits with variable coupling inductance $L_c$, the dimensionless gain-loss strength is given by $\Gamma=\gamma/\omega_0=\sqrt{L/CR^2}$. The threshold coupling at which the {\it memory-less} $\pt$-symmetric dimer transitions from $\pt$-broken region at $\mu=0$ to a $\pt$-symmetric region is given by
\begin{equation}
\label{eq:mupt}
\mu_{PT}=\sqrt{L/L_{c\mathrm{PT}}}=\sqrt{\Gamma(\Gamma+2)/2}.
\end{equation}
When the meminductance value changes from maximum to minimum, the coupling increases potentially pushing the system into the $\pt$ symmetric region; on the flip side, the reduced coupling may push the system into the     $\pt$-broken region. We investigate the behavior of the system for different meminductor strengths. The dimensionless gain-loss strength in each circuit is $\Gamma=0.5$, the initial magnetic-core fraction is given by $y(0)=0.5$, and the initial state vector $|\phi(0)\rangle$ is given by
\begin{equation}
\label{eq:chi}
|\chi_1\rangle=[0,0,0,0,i_0]^T.
\end{equation}
Note that we specify $|\phi(0)\rangle$ instead of the energy-density state vector $|\psi(0)\rangle$ because the latter also depends upon $y(0)$, i.e. the initial coupling meminductor value. 

\begin{figure*}
\centering
\includegraphics[width=0.49\textwidth]{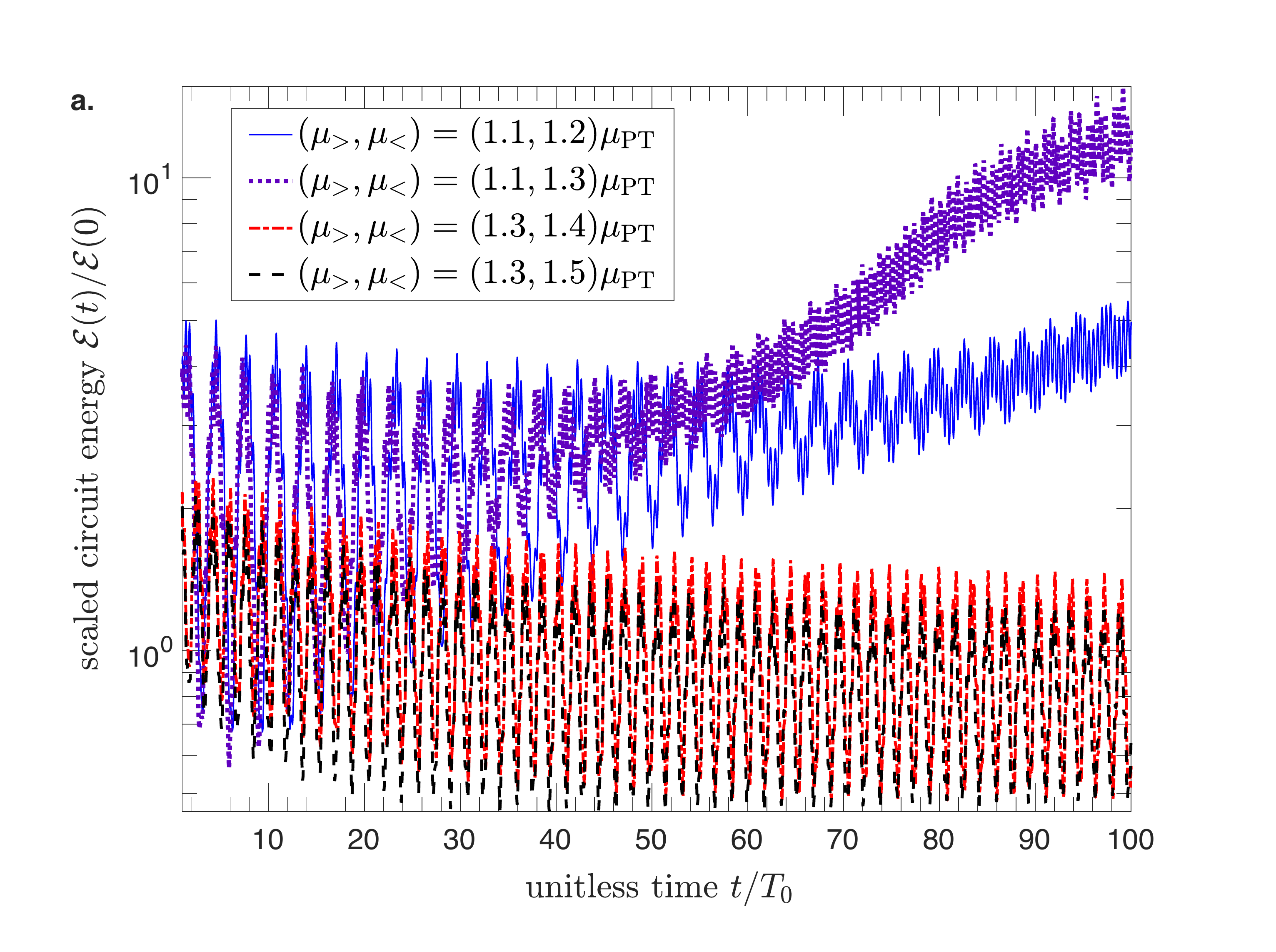}
\includegraphics[width=0.49\textwidth]{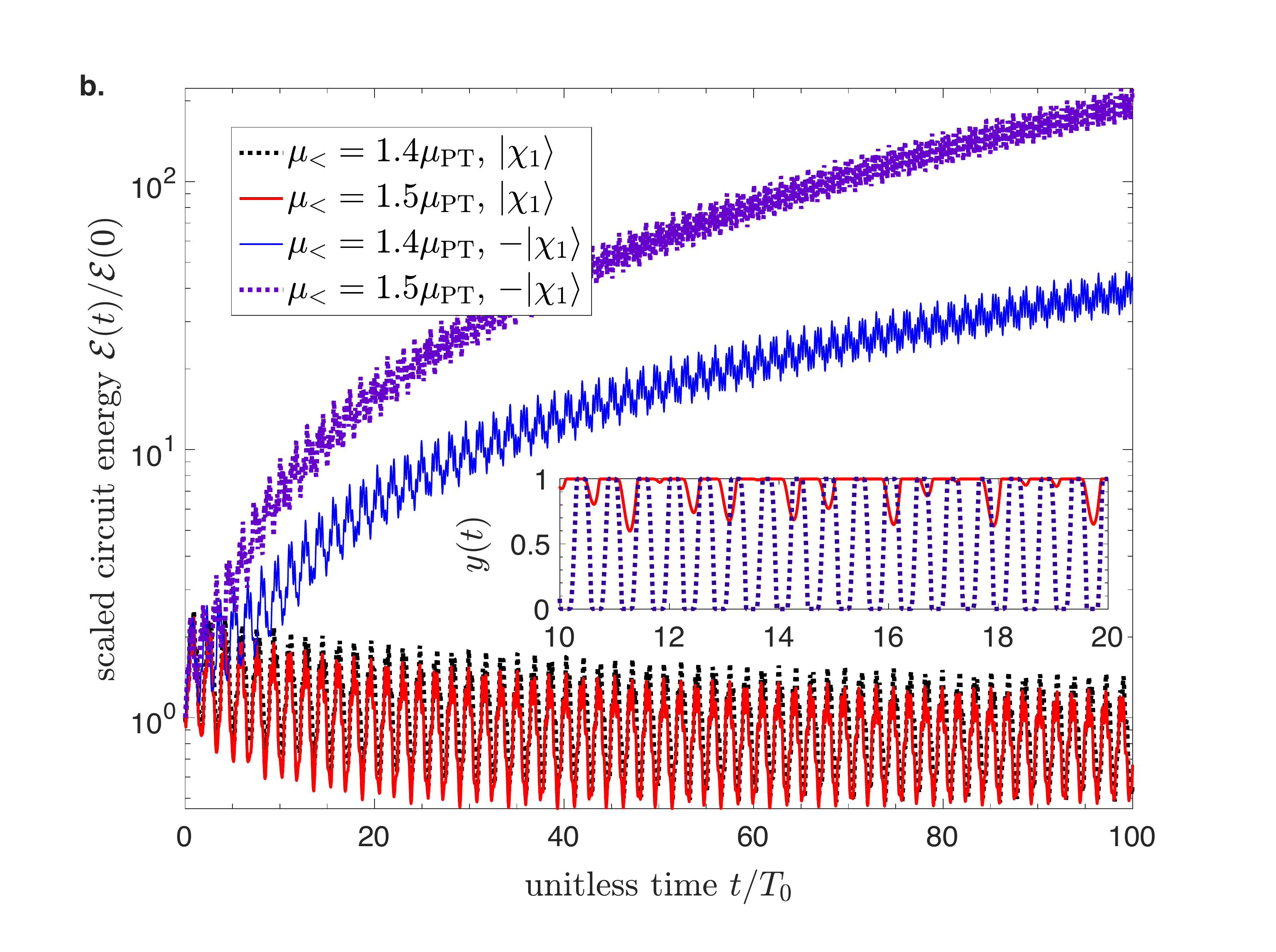}
\includegraphics[width=0.49\textwidth]{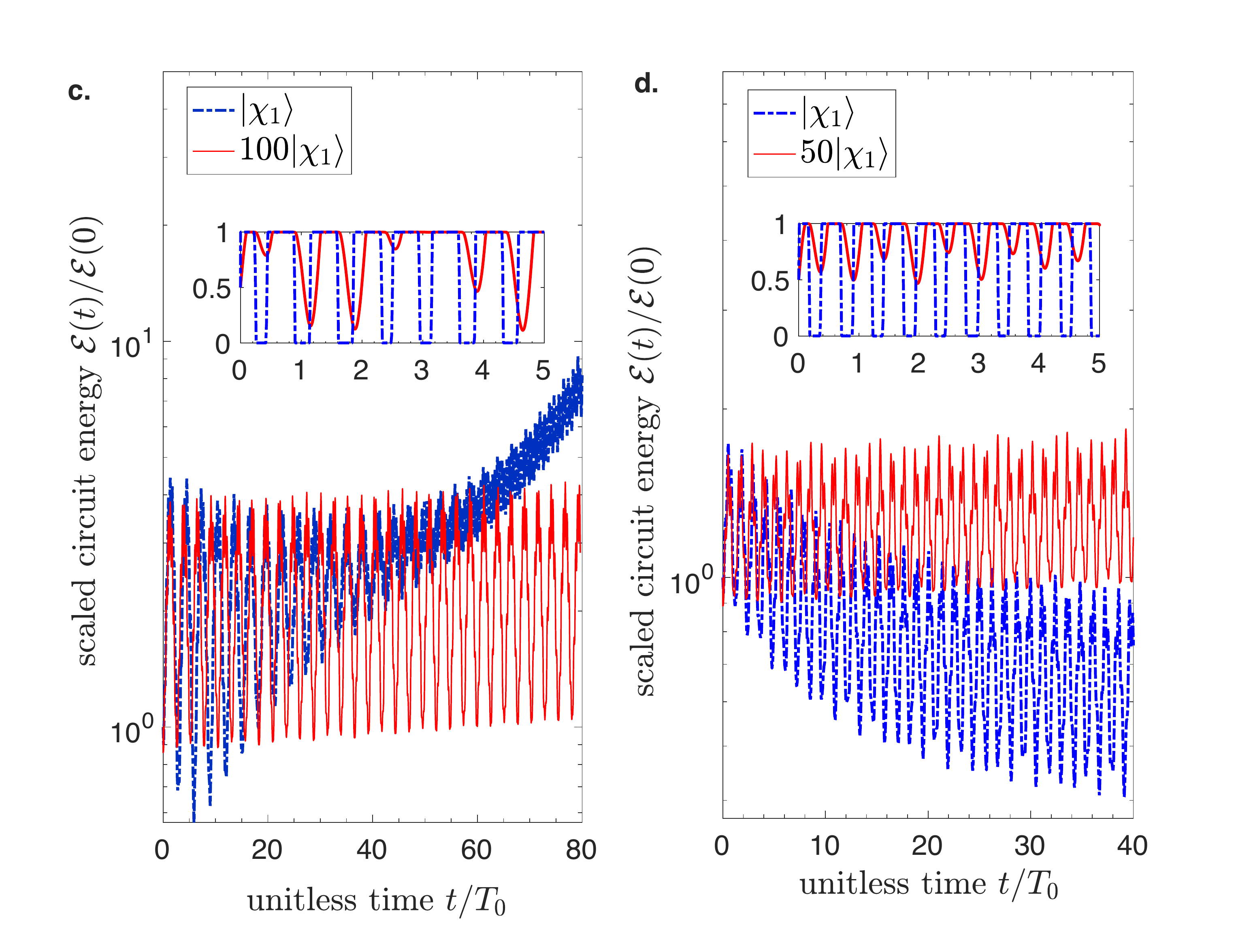}
\includegraphics[width=0.49\textwidth]{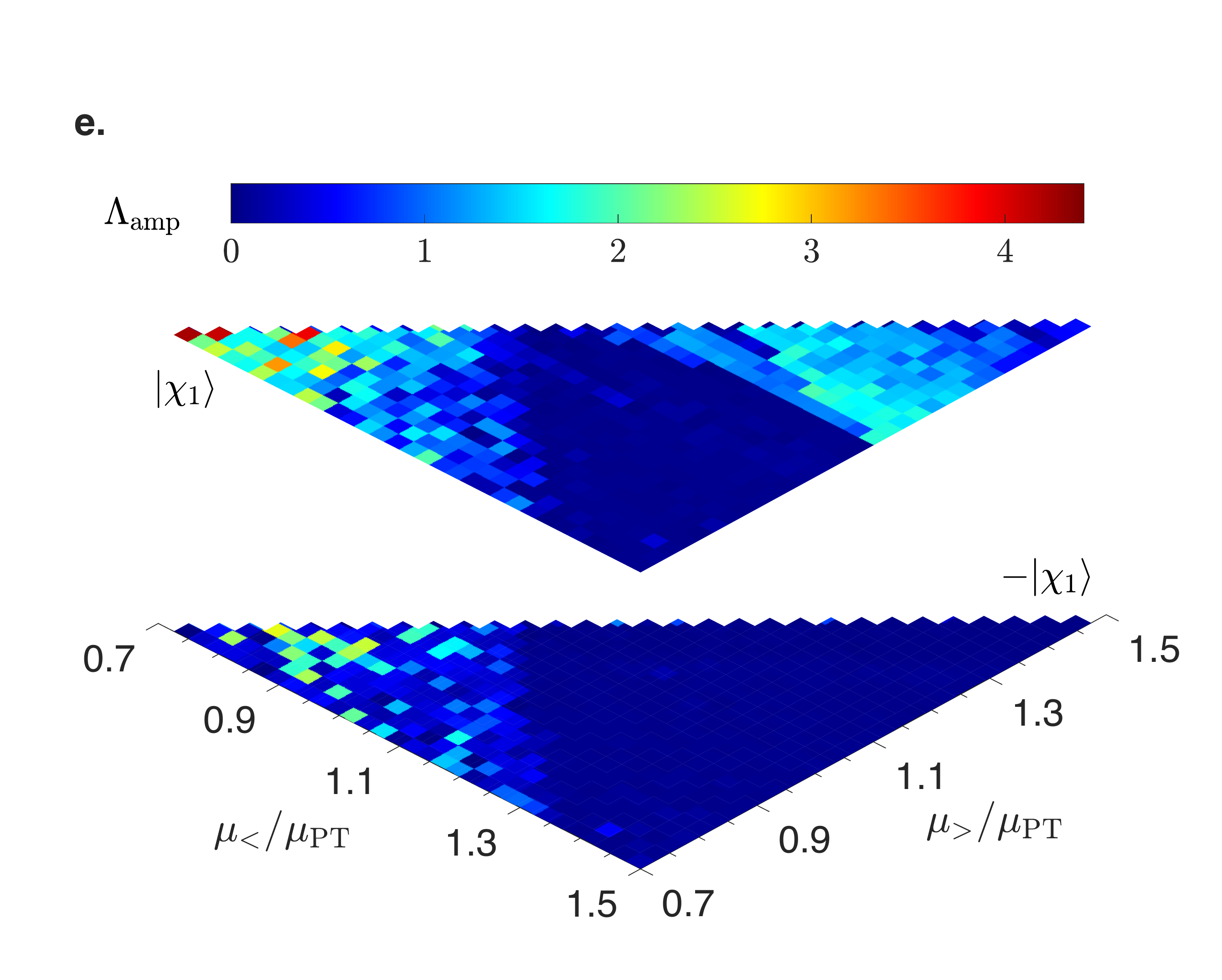}
\caption{Dynamics of a meminductive, $\pt$-symmetric dimer. (a) Scaled circuit energy $\mathcal{E}(t)/\mathcal{E}(0)$ shows transition from an exponential growth at small $\mu_<=\{1.2,1.3\}\mu_\mathrm{PT}$ to bounded oscillatory behavior at large $\mu_<=\{1.4,1.5\}\mu_\mathrm{PT}$; $\Gamma=0.5$, $y(0)=0.5$, and initial state is $|\chi_1\rangle$. (b) With initial state $-|\chi_1\rangle$, the scaled circuit energy dynamics changes from bounded oscillatory behavior to an exponential growth; $\Gamma=0.5$, $y(0)=0.5$, and $\mu_>=1.3\mu_\mathrm{PT}$. Inset: $y(t)$ oscillates in a square-wave fashion between its extremum values in the $\pt$-broken phase, whereas it does not fall below 0.5 in the $\pt$-symmetric state. The fast oscillations with period $T_0=2\pi/\omega_0$ on top of the slow dynamics are present in both phases. (c) Exponential growth of the scaled energy for initial state $|\chi_1\rangle$ is stabilized to oscillatory behavior when the initial state is changed to $100|\chi_1\rangle$, i.e. the initial meminductor current is increased 100-fold; $\Gamma=0.5$, $y(0)=0.5$, and $(\mu_>,\mu_<)=(1.1,1.3)\mu_\mathrm{PT}$. (d) For $(\mu_>,\mu_<)=(1.5,1.7)\mu_\mathrm{PT}$, and initial state $|\chi_1\rangle$, the circuit energy decays before settling into bounded, oscillatory behavior. This decay is arrested for an initial state $50\chi_1\rangle$. The insets in (c)-(d) show snapshots of the $y(t)$ dynamics. (e) Amplification factor $\Lambda_\mathrm{amp}(\mu_>,\mu_<)$ in the lower-half plane ($\mu_<$ greater than $\mu_>$) for initial states $\pm|\chi\rangle$ shows differences seen in (b).}
\label{fig:memind1}
\vspace{-5mm}
\end{figure*}

Figure~\ref{fig:memind1}a shows the transition from a $\pt$-broken phase to a $\pt$-symmetric phase that occurs when the coupling strength is increased; the vertical axis is logarithmic. Figure~\ref{fig:memind1}b shows results for $\mu_>=1.3\mu_\mathrm{PT}$ and $\mu_<=\{1.4,1.5\}\mu_\mathrm{PT}$. When the initial state is $|\chi_1\rangle$, as is seen in Fig.~\ref{fig:memind1}a, the system is in the $\pt$-symmetric phase with bounded oscillations for the scaled circuit energy. In contrast, when the initial meminductor current is reversed, i.e the initial state is given by $-|\chi_1\rangle$, the system goes into a $\pt$-symmetry broken state with exponential growth for the scaled circuit energy. The inset shows dynamics of the internal state variable $y(t)$. The key difference between oscillatory behavior and exponential growth is that for the latter, $y(t)$ switches between the two extrema in an almost square-wave fashion. Although we have shown only two instances of this highly unusual behavior, it is generically found over wide parameter ranges. It is solely due to the internal dynamics, or memory, of the coupling inductor that the fate of the system depends on the {\it sign} of the initial state, with states $\pm|\chi_1\rangle$ leading to $\pt$-symmetric and $\pt$-broken states, respectively. 

Figure~\ref{fig:memind1}c shows a typical instance where the circuit energy dynamics are stabilized by increasing the initial meminductor current. The circuit parameters are $\Gamma=0.5$, $y(0)=0.5$, and $(\mu_>,\mu_<)=(1.1,1.3)\mu_\mathrm{PT}$. We see that the exponential growth for the initial state $|\chi_1\rangle$ changes to an oscillatory energy dynamics for the state $100|\chi_1\rangle$. This qualitative change in the dynamics is reflected in the $y(t)$ dynamics (inset), where a square-wave modulation between extremum values corresponds to the $\pt$-symmetry broken state whereas oscillations that do not reach both extrema show $\pt$-symmetric phase. Another example of stabilization is shown in Fig.~\ref{fig:memind1}d; here, the circuit parameters are $(\mu_>,\mu_<)=(1.5,1.7)\mu_\mathrm{PT}$. For initial state $|\chi_1\rangle$, the scaled circuit energy decays before settling into a constant-amplitude oscillatory behavior. When the initial state is changed to $50|\chi_1\rangle$, that decay is arrested and the system settles into an oscillatory dynamics. The inset shows that both amplifying and decaying cases have internal-state variable $y(t)$ that switches between the two extrema. 

Lastly, we obtain the landscape of $\pt$-symmetric and $\pt$-broken phases via the amplification factor $\Lambda_\mathrm{amp}$ in the $\mu_<-\mu_>$ plane for initial states $\pm|\chi_1\rangle$. The relevant circuit parameters are $\Gamma=0.5$ and $y(0)=0.5$. The top plane in Fig.~\ref{fig:memind1}e shows that at small couplings $\mu_{<,>}/\mu_\mathrm{PT}\lesssim 1$, the system is in the $\pt$-broken phase. As both couplings are increased, emergence of $\pt$-symmetric phase is signaled by $\Lambda_\mathrm{amp}=0$. However, this trend is not monotonic. When the weakest coupling $\mu_>$ exceeds $\mu_\mathrm{PT}$, Eq.(\ref{eq:mupt}), the system shows a re-entrant $\pt$-symmetry broken phase. The bottom plane shows corresponding results for a system with initial state $-|\chi_1\rangle$. We see that in the region $\mu_{<,>}/\mu_\mathrm{PT}\lesssim 1$, changing the sign of the initial meminductor current does not change the fate of the system. On the other hand, some regions at moderate coupling values exhibit a change from exponential to oscillatory circuit energy dynamics. Since these results depend upon the internal state-variable value $y(0)$ and the initial circuit energy, Fig.~\ref{fig:memind1}e provides only a glimpse of the rich and diverse amplification-factor landscape for a meminductive $\pt$-symmetric dimer. 

The results in this section, too, open up many more questions than they answer. The nonlinear, initial-state sign and strength dependent dynamics of such a system, combined with the multi-dimensional relevant parameter space, make it hard to obtain significant analytical results or straightforward insights into the long-term dynamics. Our results suggest that a dynamical systems theory approach may be required to understand the long-term temporal behavior of $\pt$-symmetric systems with memory.

\section{Discussion}
\label{sec:disc}
During the past decade, open, classical systems with balanced gain and loss have seen an explosion of interest. This interest has been driven by their counterintuitive behavior, their diverse experimental realizations that span decades in relevant space and time scales, and the rich landscape of their properties that emerges when simple, canonical, static models are generalized to include time-periodicity (Floquet), time delay, noise, correlations, and nonlinearities. Here, we have presented a new paradigm for $\pt$ symmetric systems. By introducing memory in a physically meaningful and achievable manner, we have investigated the dynamics of a $\pt$-symmetric electric dimer. The nonlinearity introduced by the internal state variable that instills the memory, we find, leads to circuit energy dynamics that depend on both the strength and the sign of the initial state. Surprisingly, a $\pt$-symmetric electric dimer with memristive gain and loss shows self-organized Floquet dynamics that lead to a $\pt$-broken phase at small gain-loss strength and large coupling. Similar results are obtained when the coupling between the gain and loss $LC$ circuits have memory. It is worthwhile to point out that the energy dynamics' sensitivity to the {\it sign} of the initial state (Fig.~\ref{fig:memind1}b,e) is most unusual. Typical models for the Schrodinger equation have nonlinearities that depend on the absolute value of elements in the state vector (or the wave function). 

Our results expand the pool of simple, canonical, $\pt$-symmetric models. The notion of memory or non-Markovian behavior arises across diverse platforms, both classical and quantum. In particular, some non-Markovian aspects of quantum information and its flow between the system and its environment have been studied in dissipative~\cite{Haseli2014} or $\pt$-symmetric quantum models with a static Hamiltonian~\cite{Kawabata2017,LXiao2019,ZBian2020}. Our work, on the other hand, introduces memory into the effective non-Hermitian Hamiltonian, and leads to nonlinear and sign-dependent effects that are absent in aforementioned works. Our memory mechanism is implemented through the internal-state-variable dynamics, a non-local-in-time memory kernel, or a process tensor. Our work suggests that adding the non-Markovian aspect to $\pt$-symmetric systems will lead to non-trivial, unanticipated results.

\begin{acknowledgments}
This work was supported by IUPUI Undergraduate Research Opportunity Program (Z.A.C.), an NSF grant DMR-1054020 (Y.N.J.), and, in part, by the U.S. Department of Energy (A.S.).
\end{acknowledgments}


\bibliography{PTmemristor}

\begin{thebibliography}{56}
\expandafter\ifx\csname natexlab\endcsname\relax\def\natexlab#1{#1}\fi
\expandafter\ifx\csname bibnamefont\endcsname\relax
  \def\bibnamefont#1{#1}\fi
\expandafter\ifx\csname bibfnamefont\endcsname\relax
  \def\bibfnamefont#1{#1}\fi
\expandafter\ifx\csname citenamefont\endcsname\relax
  \def\citenamefont#1{#1}\fi
\expandafter\ifx\csname url\endcsname\relax
  \def\url#1{\texttt{#1}}\fi
\expandafter\ifx\csname urlprefix\endcsname\relax\def\urlprefix{URL }\fi
\providecommand{\bibinfo}[2]{#2}
\providecommand{\eprint}[2][]{\url{#2}}

\bibitem[{\citenamefont{Bender}(2007)}]{Bender2007}
\bibinfo{author}{\bibfnamefont{C.~M.} \bibnamefont{Bender}},
  \bibinfo{journal}{Reports on Progress in Physics}
  \textbf{\bibinfo{volume}{70}}, \bibinfo{pages}{947} (\bibinfo{year}{2007}),
  \urlprefix\url{https://doi.org/10.1088/0034-4885/70/6/r03}.

\bibitem[{\citenamefont{Bender and Boettcher}(1998)}]{Bender1998}
\bibinfo{author}{\bibfnamefont{C.~M.} \bibnamefont{Bender}} \bibnamefont{and}
  \bibinfo{author}{\bibfnamefont{S.}~\bibnamefont{Boettcher}},
  \bibinfo{journal}{Physical Review Letters} \textbf{\bibinfo{volume}{80}},
  \bibinfo{pages}{5243} (\bibinfo{year}{1998}),
  \urlprefix\url{https://doi.org/10.1103/physrevlett.80.5243}.

\bibitem[{\citenamefont{Bender et~al.}(2002)\citenamefont{Bender, Brody, and
  Jones}}]{Bender2002}
\bibinfo{author}{\bibfnamefont{C.~M.} \bibnamefont{Bender}},
  \bibinfo{author}{\bibfnamefont{D.~C.} \bibnamefont{Brody}}, \bibnamefont{and}
  \bibinfo{author}{\bibfnamefont{H.~F.} \bibnamefont{Jones}},
  \bibinfo{journal}{Physical Review Letters} \textbf{\bibinfo{volume}{89}}
  (\bibinfo{year}{2002}),
  \urlprefix\url{https://doi.org/10.1103/physrevlett.89.270401}.

\bibitem[{\citenamefont{Mostafazadeh}(2002)}]{Mostafazadeh2002}
\bibinfo{author}{\bibfnamefont{A.}~\bibnamefont{Mostafazadeh}},
  \bibinfo{journal}{Journal of Mathematical Physics}
  \textbf{\bibinfo{volume}{43}}, \bibinfo{pages}{205} (\bibinfo{year}{2002}),
  \urlprefix\url{https://doi.org/10.1063/1.1418246}.

\bibitem[{\citenamefont{Mostafazadeh}(2010)}]{Mostafazadeh2010}
\bibinfo{author}{\bibfnamefont{A.}~\bibnamefont{Mostafazadeh}},
  \bibinfo{journal}{International Journal of Geometric Methods in Modern
  Physics} \textbf{\bibinfo{volume}{07}}, \bibinfo{pages}{1191}
  (\bibinfo{year}{2010}), \eprint{https://doi.org/10.1142/S0219887810004816},
  \urlprefix\url{https://doi.org/10.1142/S0219887810004816}.

\bibitem[{\citenamefont{Joglekar et~al.}(2013)\citenamefont{Joglekar, Thompson,
  Scott, and Vemuri}}]{Joglekar2013}
\bibinfo{author}{\bibfnamefont{Y.~N.} \bibnamefont{Joglekar}},
  \bibinfo{author}{\bibfnamefont{C.}~\bibnamefont{Thompson}},
  \bibinfo{author}{\bibfnamefont{D.~D.} \bibnamefont{Scott}}, \bibnamefont{and}
  \bibinfo{author}{\bibfnamefont{G.}~\bibnamefont{Vemuri}},
  \bibinfo{journal}{The European Physical Journal Applied Physics}
  \textbf{\bibinfo{volume}{63}}, \bibinfo{pages}{30001} (\bibinfo{year}{2013}),
  \urlprefix\url{https://doi.org/10.1051/epjap/2013130240}.

\bibitem[{\citenamefont{Longhi}(2017)}]{Longhi2017}
\bibinfo{author}{\bibfnamefont{S.}~\bibnamefont{Longhi}},
  \bibinfo{journal}{{EPL} (Europhysics Letters)}
  \textbf{\bibinfo{volume}{120}}, \bibinfo{pages}{64001}
  (\bibinfo{year}{2017}),
  \urlprefix\url{https://doi.org/10.1209%2F0295-5075%2F120%2F64001}.

\bibitem[{\citenamefont{Feng et~al.}(2017)\citenamefont{Feng, El-Ganainy, and
  Ge}}]{LFeng2017}
\bibinfo{author}{\bibfnamefont{L.}~\bibnamefont{Feng}},
  \bibinfo{author}{\bibfnamefont{R.}~\bibnamefont{El-Ganainy}},
  \bibnamefont{and} \bibinfo{author}{\bibfnamefont{L.}~\bibnamefont{Ge}},
  \bibinfo{journal}{Nature Photonics} \textbf{\bibinfo{volume}{11}},
  \bibinfo{pages}{752} (\bibinfo{year}{2017}), ISSN \bibinfo{issn}{1749-4893},
  \urlprefix\url{https://doi.org/10.1038/s41566-017-0031-1}.

\bibitem[{\citenamefont{El-Ganainy et~al.}(2018)\citenamefont{El-Ganainy,
  Makris, Khajavikhan, Musslimani, Rotter, and
  Christodoulides}}]{ElGanainy2018}
\bibinfo{author}{\bibfnamefont{R.}~\bibnamefont{El-Ganainy}},
  \bibinfo{author}{\bibfnamefont{K.~G.} \bibnamefont{Makris}},
  \bibinfo{author}{\bibfnamefont{M.}~\bibnamefont{Khajavikhan}},
  \bibinfo{author}{\bibfnamefont{Z.~H.} \bibnamefont{Musslimani}},
  \bibinfo{author}{\bibfnamefont{S.}~\bibnamefont{Rotter}}, \bibnamefont{and}
  \bibinfo{author}{\bibfnamefont{D.~N.} \bibnamefont{Christodoulides}},
  \bibinfo{journal}{Nature Physics} \textbf{\bibinfo{volume}{14}},
  \bibinfo{pages}{11} (\bibinfo{year}{2018}),
  \urlprefix\url{https://doi.org/10.1038/nphys4323}.

\bibitem[{\citenamefont{Kato}(1995)}]{Kato1995}
\bibinfo{author}{\bibfnamefont{T.}~\bibnamefont{Kato}},
  \emph{\bibinfo{title}{Perturbation Theory for Linear Operators}}
  (\bibinfo{publisher}{Springer Berlin Heidelberg}, \bibinfo{year}{1995}),
  \urlprefix\url{https://doi.org/10.1007/978-3-642-66282-9}.

\bibitem[{\citenamefont{Miri and Al{\`u}}(2019)}]{miri2019}
\bibinfo{author}{\bibfnamefont{M.-A.} \bibnamefont{Miri}} \bibnamefont{and}
  \bibinfo{author}{\bibfnamefont{A.}~\bibnamefont{Al{\`u}}},
  \bibinfo{journal}{Science} \textbf{\bibinfo{volume}{363}}
  (\bibinfo{year}{2019}), ISSN \bibinfo{issn}{0036-8075},
  \urlprefix\url{https://science.sciencemag.org/content/363/6422/eaar7709}.

\bibitem[{\citenamefont{{\"O}zdemir et~al.}(2019)\citenamefont{{\"O}zdemir,
  Rotter, Nori, and Yang}}]{Ozdemir2019}
\bibinfo{author}{\bibfnamefont{S.~K.} \bibnamefont{{\"O}zdemir}},
  \bibinfo{author}{\bibfnamefont{S.}~\bibnamefont{Rotter}},
  \bibinfo{author}{\bibfnamefont{F.}~\bibnamefont{Nori}}, \bibnamefont{and}
  \bibinfo{author}{\bibfnamefont{L.}~\bibnamefont{Yang}},
  \bibinfo{journal}{Nature Materials} \textbf{\bibinfo{volume}{18}},
  \bibinfo{pages}{783} (\bibinfo{year}{2019}), ISSN \bibinfo{issn}{1476-4660},
  \urlprefix\url{https://doi.org/10.1038/s41563-019-0304-9}.

\bibitem[{\citenamefont{R\"{u}ter et~al.}(2010)\citenamefont{R\"{u}ter, Makris,
  El-Ganainy, Christodoulides, Segev, and Kip}}]{Ruter2010}
\bibinfo{author}{\bibfnamefont{C.~E.} \bibnamefont{R\"{u}ter}},
  \bibinfo{author}{\bibfnamefont{K.~G.} \bibnamefont{Makris}},
  \bibinfo{author}{\bibfnamefont{R.}~\bibnamefont{El-Ganainy}},
  \bibinfo{author}{\bibfnamefont{D.~N.} \bibnamefont{Christodoulides}},
  \bibinfo{author}{\bibfnamefont{M.}~\bibnamefont{Segev}}, \bibnamefont{and}
  \bibinfo{author}{\bibfnamefont{D.}~\bibnamefont{Kip}},
  \bibinfo{journal}{Nature Physics} \textbf{\bibinfo{volume}{6}},
  \bibinfo{pages}{192} (\bibinfo{year}{2010}),
  \urlprefix\url{https://doi.org/10.1038/nphys1515}.

\bibitem[{\citenamefont{Regensburger et~al.}(2012)\citenamefont{Regensburger,
  Bersch, Miri, Onishchukov, Christodoulides, and Peschel}}]{Regen2012}
\bibinfo{author}{\bibfnamefont{A.}~\bibnamefont{Regensburger}},
  \bibinfo{author}{\bibfnamefont{C.}~\bibnamefont{Bersch}},
  \bibinfo{author}{\bibfnamefont{M.-A.} \bibnamefont{Miri}},
  \bibinfo{author}{\bibfnamefont{G.}~\bibnamefont{Onishchukov}},
  \bibinfo{author}{\bibfnamefont{D.~N.} \bibnamefont{Christodoulides}},
  \bibnamefont{and} \bibinfo{author}{\bibfnamefont{U.}~\bibnamefont{Peschel}},
  \bibinfo{journal}{Nature} \textbf{\bibinfo{volume}{488}},
  \bibinfo{pages}{167} (\bibinfo{year}{2012}),
  \urlprefix\url{https://doi.org/10.1038/nature11298}.

\bibitem[{\citenamefont{Peng et~al.}(2014)\citenamefont{Peng, \"{O}zdemir, Lei,
  Monifi, Gianfreda, Long, Fan, Nori, Bender, and Yang}}]{Peng2014}
\bibinfo{author}{\bibfnamefont{B.}~\bibnamefont{Peng}},
  \bibinfo{author}{\bibfnamefont{{\c{S}}.~K.} \bibnamefont{\"{O}zdemir}},
  \bibinfo{author}{\bibfnamefont{F.}~\bibnamefont{Lei}},
  \bibinfo{author}{\bibfnamefont{F.}~\bibnamefont{Monifi}},
  \bibinfo{author}{\bibfnamefont{M.}~\bibnamefont{Gianfreda}},
  \bibinfo{author}{\bibfnamefont{G.~L.} \bibnamefont{Long}},
  \bibinfo{author}{\bibfnamefont{S.}~\bibnamefont{Fan}},
  \bibinfo{author}{\bibfnamefont{F.}~\bibnamefont{Nori}},
  \bibinfo{author}{\bibfnamefont{C.~M.} \bibnamefont{Bender}},
  \bibnamefont{and} \bibinfo{author}{\bibfnamefont{L.}~\bibnamefont{Yang}},
  \bibinfo{journal}{Nature Physics} \textbf{\bibinfo{volume}{10}},
  \bibinfo{pages}{394} (\bibinfo{year}{2014}),
  \urlprefix\url{https://doi.org/10.1038/nphys2927}.

\bibitem[{\citenamefont{Zhu et~al.}(2014)\citenamefont{Zhu, Ramezani, Shi, Zhu,
  and Zhang}}]{Zhu2014}
\bibinfo{author}{\bibfnamefont{X.}~\bibnamefont{Zhu}},
  \bibinfo{author}{\bibfnamefont{H.}~\bibnamefont{Ramezani}},
  \bibinfo{author}{\bibfnamefont{C.}~\bibnamefont{Shi}},
  \bibinfo{author}{\bibfnamefont{J.}~\bibnamefont{Zhu}}, \bibnamefont{and}
  \bibinfo{author}{\bibfnamefont{X.}~\bibnamefont{Zhang}},
  \bibinfo{journal}{Phys. Rev. X} \textbf{\bibinfo{volume}{4}},
  \bibinfo{pages}{031042} (\bibinfo{year}{2014}),
  \urlprefix\url{https://link.aps.org/doi/10.1103/PhysRevX.4.031042}.

\bibitem[{\citenamefont{Bender et~al.}(2013)\citenamefont{Bender, Berntson,
  Parker, and Samuel}}]{Bender2013}
\bibinfo{author}{\bibfnamefont{C.~M.} \bibnamefont{Bender}},
  \bibinfo{author}{\bibfnamefont{B.~K.} \bibnamefont{Berntson}},
  \bibinfo{author}{\bibfnamefont{D.}~\bibnamefont{Parker}}, \bibnamefont{and}
  \bibinfo{author}{\bibfnamefont{E.}~\bibnamefont{Samuel}},
  \bibinfo{journal}{American Journal of Physics} \textbf{\bibinfo{volume}{81}},
  \bibinfo{pages}{173} (\bibinfo{year}{2013}),
  \urlprefix\url{https://doi.org/10.1119/1.4789549}.

\bibitem[{\citenamefont{Schindler et~al.}(2011)\citenamefont{Schindler, Li,
  Zheng, Ellis, and Kottos}}]{Schindler2011}
\bibinfo{author}{\bibfnamefont{J.}~\bibnamefont{Schindler}},
  \bibinfo{author}{\bibfnamefont{A.}~\bibnamefont{Li}},
  \bibinfo{author}{\bibfnamefont{M.~C.} \bibnamefont{Zheng}},
  \bibinfo{author}{\bibfnamefont{F.~M.} \bibnamefont{Ellis}}, \bibnamefont{and}
  \bibinfo{author}{\bibfnamefont{T.}~\bibnamefont{Kottos}},
  \bibinfo{journal}{Phys. Rev. A} \textbf{\bibinfo{volume}{84}},
  \bibinfo{pages}{040101} (\bibinfo{year}{2011}),
  \urlprefix\url{https://link.aps.org/doi/10.1103/PhysRevA.84.040101}.

\bibitem[{\citenamefont{Wang et~al.}(2020)\citenamefont{Wang, Fang, Xie, Dong,
  Joglekar, Wang, Li, and Luo}}]{Wang2020}
\bibinfo{author}{\bibfnamefont{T.}~\bibnamefont{Wang}},
  \bibinfo{author}{\bibfnamefont{J.}~\bibnamefont{Fang}},
  \bibinfo{author}{\bibfnamefont{Z.}~\bibnamefont{Xie}},
  \bibinfo{author}{\bibfnamefont{N.}~\bibnamefont{Dong}},
  \bibinfo{author}{\bibfnamefont{Y.~N.} \bibnamefont{Joglekar}},
  \bibinfo{author}{\bibfnamefont{Z.}~\bibnamefont{Wang}},
  \bibinfo{author}{\bibfnamefont{J.}~\bibnamefont{Li}}, \bibnamefont{and}
  \bibinfo{author}{\bibfnamefont{L.}~\bibnamefont{Luo}}, \bibinfo{journal}{The
  European Physical Journal D} \textbf{\bibinfo{volume}{74}}
  (\bibinfo{year}{2020}),
  \urlprefix\url{https://doi.org/10.1140/epjd/e2020-10131-7}.

\bibitem[{\citenamefont{Caves}(1982)}]{Caves1982}
\bibinfo{author}{\bibfnamefont{C.~M.} \bibnamefont{Caves}},
  \bibinfo{journal}{Phys. Rev. D} \textbf{\bibinfo{volume}{26}},
  \bibinfo{pages}{1817} (\bibinfo{year}{1982}),
  \urlprefix\url{https://link.aps.org/doi/10.1103/PhysRevD.26.1817}.

\bibitem[{\citenamefont{Scheel and Szameit}(2018)}]{Scheel2018}
\bibinfo{author}{\bibfnamefont{S.}~\bibnamefont{Scheel}} \bibnamefont{and}
  \bibinfo{author}{\bibfnamefont{A.}~\bibnamefont{Szameit}},
  \bibinfo{journal}{{EPL} (Europhysics Letters)}
  \textbf{\bibinfo{volume}{122}}, \bibinfo{pages}{34001}
  (\bibinfo{year}{2018}),
  \urlprefix\url{https://doi.org/10.1209/0295-5075/122/34001}.

\bibitem[{\citenamefont{Guo et~al.}(2009)\citenamefont{Guo, Salamo, Duchesne,
  Morandotti, Volatier-Ravat, Aimez, Siviloglou, and
  Christodoulides}}]{Guo2009}
\bibinfo{author}{\bibfnamefont{A.}~\bibnamefont{Guo}},
  \bibinfo{author}{\bibfnamefont{G.~J.} \bibnamefont{Salamo}},
  \bibinfo{author}{\bibfnamefont{D.}~\bibnamefont{Duchesne}},
  \bibinfo{author}{\bibfnamefont{R.}~\bibnamefont{Morandotti}},
  \bibinfo{author}{\bibfnamefont{M.}~\bibnamefont{Volatier-Ravat}},
  \bibinfo{author}{\bibfnamefont{V.}~\bibnamefont{Aimez}},
  \bibinfo{author}{\bibfnamefont{G.~A.} \bibnamefont{Siviloglou}},
  \bibnamefont{and} \bibinfo{author}{\bibfnamefont{D.~N.}
  \bibnamefont{Christodoulides}}, \bibinfo{journal}{Physical Review Letters}
  \textbf{\bibinfo{volume}{103}} (\bibinfo{year}{2009}),
  \urlprefix\url{https://doi.org/10.1103/physrevlett.103.093902}.

\bibitem[{\citenamefont{Joglekar and Harter}(2018)}]{Joglekar2018}
\bibinfo{author}{\bibfnamefont{Y.~N.} \bibnamefont{Joglekar}} \bibnamefont{and}
  \bibinfo{author}{\bibfnamefont{A.~K.} \bibnamefont{Harter}},
  \bibinfo{journal}{Photonics Research} \textbf{\bibinfo{volume}{6}},
  \bibinfo{pages}{A51} (\bibinfo{year}{2018}),
  \urlprefix\url{https://doi.org/10.1364/prj.6.000a51}.

\bibitem[{\citenamefont{Klauck et~al.}(2019)\citenamefont{Klauck, Teuber,
  Ornigotti, Heinrich, Scheel, and Szameit}}]{Klauck2019}
\bibinfo{author}{\bibfnamefont{F.}~\bibnamefont{Klauck}},
  \bibinfo{author}{\bibfnamefont{L.}~\bibnamefont{Teuber}},
  \bibinfo{author}{\bibfnamefont{M.}~\bibnamefont{Ornigotti}},
  \bibinfo{author}{\bibfnamefont{M.}~\bibnamefont{Heinrich}},
  \bibinfo{author}{\bibfnamefont{S.}~\bibnamefont{Scheel}}, \bibnamefont{and}
  \bibinfo{author}{\bibfnamefont{A.}~\bibnamefont{Szameit}},
  \bibinfo{journal}{Nature Photonics} \textbf{\bibinfo{volume}{13}},
  \bibinfo{pages}{883} (\bibinfo{year}{2019}),
  \urlprefix\url{https://doi.org/10.1038/s41566-019-0517-0}.

\bibitem[{\citenamefont{Li et~al.}(2019)\citenamefont{Li, Harter, Liu, de~Melo,
  Joglekar, and Luo}}]{Li2019a}
\bibinfo{author}{\bibfnamefont{J.}~\bibnamefont{Li}},
  \bibinfo{author}{\bibfnamefont{A.~K.} \bibnamefont{Harter}},
  \bibinfo{author}{\bibfnamefont{J.}~\bibnamefont{Liu}},
  \bibinfo{author}{\bibfnamefont{L.}~\bibnamefont{de~Melo}},
  \bibinfo{author}{\bibfnamefont{Y.~N.} \bibnamefont{Joglekar}},
  \bibnamefont{and} \bibinfo{author}{\bibfnamefont{L.}~\bibnamefont{Luo}},
  \bibinfo{journal}{Nature Communications} \textbf{\bibinfo{volume}{10}}
  (\bibinfo{year}{2019}),
  \urlprefix\url{https://doi.org/10.1038/s41467-019-08596-1}.

\bibitem[{\citenamefont{Wu et~al.}(2019)\citenamefont{Wu, Liu, Geng, Song, Ye,
  Duan, Rong, and Du}}]{Wu2019}
\bibinfo{author}{\bibfnamefont{Y.}~\bibnamefont{Wu}},
  \bibinfo{author}{\bibfnamefont{W.}~\bibnamefont{Liu}},
  \bibinfo{author}{\bibfnamefont{J.}~\bibnamefont{Geng}},
  \bibinfo{author}{\bibfnamefont{X.}~\bibnamefont{Song}},
  \bibinfo{author}{\bibfnamefont{X.}~\bibnamefont{Ye}},
  \bibinfo{author}{\bibfnamefont{C.-K.} \bibnamefont{Duan}},
  \bibinfo{author}{\bibfnamefont{X.}~\bibnamefont{Rong}}, \bibnamefont{and}
  \bibinfo{author}{\bibfnamefont{J.}~\bibnamefont{Du}},
  \bibinfo{journal}{Science} \textbf{\bibinfo{volume}{364}},
  \bibinfo{pages}{878} (\bibinfo{year}{2019}),
  \urlprefix\url{https://doi.org/10.1126/science.aaw8205}.

\bibitem[{\citenamefont{Naghiloo et~al.}(2019)\citenamefont{Naghiloo, Abbasi,
  Joglekar, and Murch}}]{Naghiloo2019}
\bibinfo{author}{\bibfnamefont{M.}~\bibnamefont{Naghiloo}},
  \bibinfo{author}{\bibfnamefont{M.}~\bibnamefont{Abbasi}},
  \bibinfo{author}{\bibfnamefont{Y.~N.} \bibnamefont{Joglekar}},
  \bibnamefont{and} \bibinfo{author}{\bibfnamefont{K.~W.} \bibnamefont{Murch}},
  \bibinfo{journal}{Nature Physics} \textbf{\bibinfo{volume}{15}},
  \bibinfo{pages}{1232} (\bibinfo{year}{2019}),
  \urlprefix\url{https://doi.org/10.1038/s41567-019-0652-z}.

\bibitem[{\citenamefont{Joglekar et~al.}(2014)\citenamefont{Joglekar, Marathe,
  Durganandini, and Pathak}}]{Joglekar2014}
\bibinfo{author}{\bibfnamefont{Y.~N.} \bibnamefont{Joglekar}},
  \bibinfo{author}{\bibfnamefont{R.}~\bibnamefont{Marathe}},
  \bibinfo{author}{\bibfnamefont{P.}~\bibnamefont{Durganandini}},
  \bibnamefont{and} \bibinfo{author}{\bibfnamefont{R.~K.}
  \bibnamefont{Pathak}}, \bibinfo{journal}{Physical Review A}
  \textbf{\bibinfo{volume}{90}} (\bibinfo{year}{2014}),
  \urlprefix\url{https://doi.org/10.1103/physreva.90.040101}.

\bibitem[{\citenamefont{Lee and Joglekar}(2015)}]{Lee2015}
\bibinfo{author}{\bibfnamefont{T.~E.} \bibnamefont{Lee}} \bibnamefont{and}
  \bibinfo{author}{\bibfnamefont{Y.~N.} \bibnamefont{Joglekar}},
  \bibinfo{journal}{Physical Review A} \textbf{\bibinfo{volume}{92}}
  (\bibinfo{year}{2015}),
  \urlprefix\url{https://doi.org/10.1103/physreva.92.042103}.

\bibitem[{\citenamefont{Chitsazi et~al.}(2017)\citenamefont{Chitsazi, Li,
  Ellis, and Kottos}}]{Chitsazi2017}
\bibinfo{author}{\bibfnamefont{M.}~\bibnamefont{Chitsazi}},
  \bibinfo{author}{\bibfnamefont{H.}~\bibnamefont{Li}},
  \bibinfo{author}{\bibfnamefont{F.}~\bibnamefont{Ellis}}, \bibnamefont{and}
  \bibinfo{author}{\bibfnamefont{T.}~\bibnamefont{Kottos}},
  \bibinfo{journal}{Physical Review Letters} \textbf{\bibinfo{volume}{119}}
  (\bibinfo{year}{2017}),
  \urlprefix\url{https://doi.org/10.1103/physrevlett.119.093901}.

\bibitem[{\citenamefont{de~J.~Le{\'{o}}n-Montiel
  et~al.}(2018)\citenamefont{de~J.~Le{\'{o}}n-Montiel, Quiroz-Ju{\'{a}}rez,
  Dom{\'{\i}}nguez-Ju{\'{a}}rez, Quintero-Torres, Arag{\'{o}}n, Harter, and
  Joglekar}}]{LeonMontiel2018}
\bibinfo{author}{\bibfnamefont{R.}~\bibnamefont{de~J.~Le{\'{o}}n-Montiel}},
  \bibinfo{author}{\bibfnamefont{M.~A.} \bibnamefont{Quiroz-Ju{\'{a}}rez}},
  \bibinfo{author}{\bibfnamefont{J.~L.}
  \bibnamefont{Dom{\'{\i}}nguez-Ju{\'{a}}rez}},
  \bibinfo{author}{\bibfnamefont{R.}~\bibnamefont{Quintero-Torres}},
  \bibinfo{author}{\bibfnamefont{J.~L.} \bibnamefont{Arag{\'{o}}n}},
  \bibinfo{author}{\bibfnamefont{A.~K.} \bibnamefont{Harter}},
  \bibnamefont{and} \bibinfo{author}{\bibfnamefont{Y.~N.}
  \bibnamefont{Joglekar}}, \bibinfo{journal}{Communications Physics}
  \textbf{\bibinfo{volume}{1}} (\bibinfo{year}{2018}),
  \urlprefix\url{https://doi.org/10.1038/s42005-018-0087-3}.

\bibitem[{\citenamefont{Konotop et~al.}(2016)\citenamefont{Konotop, Yang, and
  Zezyulin}}]{Konotop2016}
\bibinfo{author}{\bibfnamefont{V.~V.} \bibnamefont{Konotop}},
  \bibinfo{author}{\bibfnamefont{J.}~\bibnamefont{Yang}}, \bibnamefont{and}
  \bibinfo{author}{\bibfnamefont{D.~A.} \bibnamefont{Zezyulin}},
  \bibinfo{journal}{Rev. Mod. Phys.} \textbf{\bibinfo{volume}{88}},
  \bibinfo{pages}{035002} (\bibinfo{year}{2016}),
  \urlprefix\url{https://link.aps.org/doi/10.1103/RevModPhys.88.035002}.

\bibitem[{\citenamefont{Wilkey et~al.}(2019)\citenamefont{Wilkey, Joglekar,
  Suelzer, and Vemuri}}]{Wilkey2019}
\bibinfo{author}{\bibfnamefont{A.}~\bibnamefont{Wilkey}},
  \bibinfo{author}{\bibfnamefont{Y.}~\bibnamefont{Joglekar}},
  \bibinfo{author}{\bibfnamefont{J.~S.} \bibnamefont{Suelzer}},
  \bibnamefont{and} \bibinfo{author}{\bibfnamefont{G.}~\bibnamefont{Vemuri}},
  in \emph{\bibinfo{booktitle}{Active Photonic Platforms XI}}, edited by
  \bibinfo{editor}{\bibfnamefont{G.~S.} \bibnamefont{Subramania}}
  \bibnamefont{and}
  \bibinfo{editor}{\bibfnamefont{S.}~\bibnamefont{Foteinopoulou}},
  \bibinfo{organization}{International Society for Optics and Photonics}
  (\bibinfo{publisher}{SPIE}, \bibinfo{year}{2019}), vol.
  \bibinfo{volume}{11081}, pp. \bibinfo{pages}{9 -- 22},
  \urlprefix\url{https://doi.org/10.1117/12.2523786}.

\bibitem[{\citenamefont{Joglekar}(2010)}]{Joglekar2010}
\bibinfo{author}{\bibfnamefont{Y.~N.} \bibnamefont{Joglekar}},
  \bibinfo{journal}{Phys. Rev. A} \textbf{\bibinfo{volume}{82}},
  \bibinfo{pages}{044101} (\bibinfo{year}{2010}),
  \urlprefix\url{https://link.aps.org/doi/10.1103/PhysRevA.82.044101}.

\bibitem[{\citenamefont{Ventra et~al.}(2009{\natexlab{a}})\citenamefont{Ventra,
  Pershin, and Chua}}]{DiVentra2009}
\bibinfo{author}{\bibfnamefont{M.~D.} \bibnamefont{Ventra}},
  \bibinfo{author}{\bibfnamefont{Y.~V.} \bibnamefont{Pershin}},
  \bibnamefont{and} \bibinfo{author}{\bibfnamefont{L.~O.} \bibnamefont{Chua}},
  \bibinfo{journal}{Proceedings of the {IEEE}} \textbf{\bibinfo{volume}{97}},
  \bibinfo{pages}{1717} (\bibinfo{year}{2009}{\natexlab{a}}),
  \urlprefix\url{https://doi.org/10.1109/jproc.2009.2021077}.

\bibitem[{\citenamefont{Chua}(1971)}]{Chua1971}
\bibinfo{author}{\bibfnamefont{L.}~\bibnamefont{Chua}},
  \bibinfo{journal}{{IEEE} Transactions on Circuit Theory}
  \textbf{\bibinfo{volume}{18}}, \bibinfo{pages}{507} (\bibinfo{year}{1971}),
  \urlprefix\url{https://doi.org/10.1109/tct.1971.1083337}.

\bibitem[{\citenamefont{Strukov et~al.}(2008)\citenamefont{Strukov, Snider,
  Stewart, and Williams}}]{Strukov2008}
\bibinfo{author}{\bibfnamefont{D.~B.} \bibnamefont{Strukov}},
  \bibinfo{author}{\bibfnamefont{G.~S.} \bibnamefont{Snider}},
  \bibinfo{author}{\bibfnamefont{D.~R.} \bibnamefont{Stewart}},
  \bibnamefont{and} \bibinfo{author}{\bibfnamefont{R.~S.}
  \bibnamefont{Williams}}, \bibinfo{journal}{Nature}
  \textbf{\bibinfo{volume}{453}}, \bibinfo{pages}{80} (\bibinfo{year}{2008}),
  \urlprefix\url{https://doi.org/10.1038/nature06932}.

\bibitem[{\citenamefont{Tour and He}(2008)}]{Tour2008}
\bibinfo{author}{\bibfnamefont{J.~M.} \bibnamefont{Tour}} \bibnamefont{and}
  \bibinfo{author}{\bibfnamefont{T.}~\bibnamefont{He}},
  \bibinfo{journal}{Nature} \textbf{\bibinfo{volume}{453}}, \bibinfo{pages}{42}
  (\bibinfo{year}{2008}), \urlprefix\url{https://doi.org/10.1038/453042a}.

\bibitem[{\citenamefont{Chua}(2013)}]{Chua2013}
\bibinfo{author}{\bibfnamefont{L.}~\bibnamefont{Chua}}, in
  \emph{\bibinfo{booktitle}{Memristors and Memristive Systems}}
  (\bibinfo{publisher}{Springer New York}, \bibinfo{year}{2013}), pp.
  \bibinfo{pages}{17--90},
  \urlprefix\url{https://doi.org/10.1007/978-1-4614-9068-5_2}.

\bibitem[{\citenamefont{Sapoff and Oppenheim}(1963)}]{Sapoff1963}
\bibinfo{author}{\bibfnamefont{M.}~\bibnamefont{Sapoff}} \bibnamefont{and}
  \bibinfo{author}{\bibfnamefont{R.}~\bibnamefont{Oppenheim}},
  \bibinfo{journal}{Proceedings of the {IEEE}} \textbf{\bibinfo{volume}{51}},
  \bibinfo{pages}{1292} (\bibinfo{year}{1963}),
  \urlprefix\url{https://doi.org/10.1109/proc.1963.2560}.

\bibitem[{\citenamefont{Hodgkin and Huxley}(1952)}]{Hodgkin1952}
\bibinfo{author}{\bibfnamefont{A.~L.} \bibnamefont{Hodgkin}} \bibnamefont{and}
  \bibinfo{author}{\bibfnamefont{A.~F.} \bibnamefont{Huxley}},
  \bibinfo{journal}{The Journal of Physiology} \textbf{\bibinfo{volume}{117}},
  \bibinfo{pages}{500} (\bibinfo{year}{1952}),
  \urlprefix\url{https://doi.org/10.1113/jphysiol.1952.sp004764}.

\bibitem[{\citenamefont{CHUA et~al.}(2012)\citenamefont{CHUA, SBITNEV, and
  KIM}}]{CHUA2012}
\bibinfo{author}{\bibfnamefont{L.}~\bibnamefont{CHUA}},
  \bibinfo{author}{\bibfnamefont{V.}~\bibnamefont{SBITNEV}}, \bibnamefont{and}
  \bibinfo{author}{\bibfnamefont{H.}~\bibnamefont{KIM}},
  \bibinfo{journal}{International Journal of Bifurcation and Chaos}
  \textbf{\bibinfo{volume}{22}}, \bibinfo{pages}{1230011}
  (\bibinfo{year}{2012}),
  \urlprefix\url{https://doi.org/10.1142/s021812741230011x}.

\bibitem[{\citenamefont{Joglekar and Wolf}(2009)}]{Joglekar2009}
\bibinfo{author}{\bibfnamefont{Y.~N.} \bibnamefont{Joglekar}} \bibnamefont{and}
  \bibinfo{author}{\bibfnamefont{S.~J.} \bibnamefont{Wolf}},
  \bibinfo{journal}{European Journal of Physics} \textbf{\bibinfo{volume}{30}},
  \bibinfo{pages}{661} (\bibinfo{year}{2009}),
  \urlprefix\url{https://doi.org/10.1088/0143-0807/30/4/001}.

\bibitem[{\citenamefont{Joglekar and Meijome}(2012)}]{Joglekar2012m}
\bibinfo{author}{\bibfnamefont{Y.~N.} \bibnamefont{Joglekar}} \bibnamefont{and}
  \bibinfo{author}{\bibfnamefont{N.}~\bibnamefont{Meijome}},
  \bibinfo{journal}{{IEEE} Transactions on Circuits and Systems {II}: Express
  Briefs} \textbf{\bibinfo{volume}{59}}, \bibinfo{pages}{830}
  (\bibinfo{year}{2012}),
  \urlprefix\url{https://doi.org/10.1109/tcsii.2012.2220692}.

\bibitem[{\citenamefont{Hofmann et~al.}(2020)\citenamefont{Hofmann, Helbig,
  Schindler, Salgo, Brzezi\ifmmode~\acute{n}\else \'{n}\fi{}ska, Greiter,
  Kiessling, Wolf, Vollhardt, Kaba\ifmmode~\check{s}\else \v{s}\fi{}i
  et~al.}}]{Hoffmann2020}
\bibinfo{author}{\bibfnamefont{T.}~\bibnamefont{Hofmann}},
  \bibinfo{author}{\bibfnamefont{T.}~\bibnamefont{Helbig}},
  \bibinfo{author}{\bibfnamefont{F.}~\bibnamefont{Schindler}},
  \bibinfo{author}{\bibfnamefont{N.}~\bibnamefont{Salgo}},
  \bibinfo{author}{\bibfnamefont{M.}~\bibnamefont{Brzezi\ifmmode~\acute{n}\else
  \'{n}\fi{}ska}}, \bibinfo{author}{\bibfnamefont{M.}~\bibnamefont{Greiter}},
  \bibinfo{author}{\bibfnamefont{T.}~\bibnamefont{Kiessling}},
  \bibinfo{author}{\bibfnamefont{D.}~\bibnamefont{Wolf}},
  \bibinfo{author}{\bibfnamefont{A.}~\bibnamefont{Vollhardt}},
  \bibinfo{author}{\bibfnamefont{A.}~\bibnamefont{Kaba\ifmmode~\check{s}\else
  \v{s}\fi{}i}}, \bibnamefont{et~al.}, \bibinfo{journal}{Phys. Rev. Research}
  \textbf{\bibinfo{volume}{2}}, \bibinfo{pages}{023265} (\bibinfo{year}{2020}),
  \urlprefix\url{https://link.aps.org/doi/10.1103/PhysRevResearch.2.023265}.

\bibitem[{\citenamefont{Kim et~al.}(2012)\citenamefont{Kim, Sah, Yang, Cho, and
  Chua}}]{HKim2012}
\bibinfo{author}{\bibfnamefont{H.}~\bibnamefont{Kim}},
  \bibinfo{author}{\bibfnamefont{M.~P.} \bibnamefont{Sah}},
  \bibinfo{author}{\bibfnamefont{C.}~\bibnamefont{Yang}},
  \bibinfo{author}{\bibfnamefont{S.}~\bibnamefont{Cho}}, \bibnamefont{and}
  \bibinfo{author}{\bibfnamefont{L.~O.} \bibnamefont{Chua}},
  \bibinfo{journal}{{IEEE} Transactions on Circuits and Systems I: Regular
  Papers} \textbf{\bibinfo{volume}{59}}, \bibinfo{pages}{2422}
  (\bibinfo{year}{2012}),
  \urlprefix\url{https://doi.org/10.1109/tcsi.2012.2188957}.

\bibitem[{\citenamefont{Ventra et~al.}(2009{\natexlab{b}})\citenamefont{Ventra,
  Pershin, and Chua}}]{Ventra2009}
\bibinfo{author}{\bibfnamefont{M.~D.} \bibnamefont{Ventra}},
  \bibinfo{author}{\bibfnamefont{Y.~V.} \bibnamefont{Pershin}},
  \bibnamefont{and} \bibinfo{author}{\bibfnamefont{L.~O.} \bibnamefont{Chua}},
  \bibinfo{journal}{Proceedings of the {IEEE}} \textbf{\bibinfo{volume}{97}},
  \bibinfo{pages}{1717} (\bibinfo{year}{2009}{\natexlab{b}}),
  \urlprefix\url{https://doi.org/10.1109/jproc.2009.2021077}.

\bibitem[{\citenamefont{Yin et~al.}(2015)\citenamefont{Yin, Tian, Chen, and
  Chua}}]{ZYin2015}
\bibinfo{author}{\bibfnamefont{Z.}~\bibnamefont{Yin}},
  \bibinfo{author}{\bibfnamefont{H.}~\bibnamefont{Tian}},
  \bibinfo{author}{\bibfnamefont{G.}~\bibnamefont{Chen}}, \bibnamefont{and}
  \bibinfo{author}{\bibfnamefont{L.~O.} \bibnamefont{Chua}},
  \bibinfo{journal}{{IEEE} Transactions on Circuits and Systems {II}: Express
  Briefs} \textbf{\bibinfo{volume}{62}}, \bibinfo{pages}{402}
  (\bibinfo{year}{2015}),
  \urlprefix\url{https://doi.org/10.1109/tcsii.2014.2387653}.

\bibitem[{\citenamefont{Slade}(2005)}]{Slade2005}
\bibinfo{author}{\bibfnamefont{G.}~\bibnamefont{Slade}},
  \bibinfo{journal}{{IEEE} Transactions on Magnetics}
  \textbf{\bibinfo{volume}{41}}, \bibinfo{pages}{4270} (\bibinfo{year}{2005}),
  \urlprefix\url{https://doi.org/10.1109/tmag.2005.856320}.

\bibitem[{\citenamefont{Dyakonov}(2007)}]{Dyakonov2007}
\bibinfo{author}{\bibfnamefont{M.~I.} \bibnamefont{Dyakonov}},
  \bibinfo{journal}{Physical Review Letters} \textbf{\bibinfo{volume}{99}}
  (\bibinfo{year}{2007}),
  \urlprefix\url{https://doi.org/10.1103/physrevlett.99.126601}.

\bibitem[{\citenamefont{Miao et~al.}(2014)\citenamefont{Miao, Huang, Qu, and
  Chien}}]{Miao2014}
\bibinfo{author}{\bibfnamefont{B.}~\bibnamefont{Miao}},
  \bibinfo{author}{\bibfnamefont{S.}~\bibnamefont{Huang}},
  \bibinfo{author}{\bibfnamefont{D.}~\bibnamefont{Qu}}, \bibnamefont{and}
  \bibinfo{author}{\bibfnamefont{C.}~\bibnamefont{Chien}},
  \bibinfo{journal}{Physical Review Letters} \textbf{\bibinfo{volume}{112}}
  (\bibinfo{year}{2014}),
  \urlprefix\url{https://doi.org/10.1103/physrevlett.112.236601}.

\bibitem[{\citenamefont{Han et~al.}(2014)\citenamefont{Han, Song, Gao, Wang,
  Chen, and Pan}}]{Han2014}
\bibinfo{author}{\bibfnamefont{J.}~\bibnamefont{Han}},
  \bibinfo{author}{\bibfnamefont{C.}~\bibnamefont{Song}},
  \bibinfo{author}{\bibfnamefont{S.}~\bibnamefont{Gao}},
  \bibinfo{author}{\bibfnamefont{Y.}~\bibnamefont{Wang}},
  \bibinfo{author}{\bibfnamefont{C.}~\bibnamefont{Chen}}, \bibnamefont{and}
  \bibinfo{author}{\bibfnamefont{F.}~\bibnamefont{Pan}},
  \bibinfo{journal}{{ACS} Nano} \textbf{\bibinfo{volume}{8}},
  \bibinfo{pages}{10043} (\bibinfo{year}{2014}),
  \urlprefix\url{https://doi.org/10.1021/nn502655u}.

\bibitem[{\citenamefont{Haseli et~al.}(2014)\citenamefont{Haseli, Karpat,
  Salimi, Khorashad, Fanchini, \ifmmode~\mbox{\c{C}}\else \c{C}\fi{}akmak,
  Aguilar, Walborn, and Ribeiro}}]{Haseli2014}
\bibinfo{author}{\bibfnamefont{S.}~\bibnamefont{Haseli}},
  \bibinfo{author}{\bibfnamefont{G.}~\bibnamefont{Karpat}},
  \bibinfo{author}{\bibfnamefont{S.}~\bibnamefont{Salimi}},
  \bibinfo{author}{\bibfnamefont{A.~S.} \bibnamefont{Khorashad}},
  \bibinfo{author}{\bibfnamefont{F.~F.} \bibnamefont{Fanchini}},
  \bibinfo{author}{\bibfnamefont{B.}~\bibnamefont{\ifmmode~\mbox{\c{C}}\else
  \c{C}\fi{}akmak}}, \bibinfo{author}{\bibfnamefont{G.~H.}
  \bibnamefont{Aguilar}}, \bibinfo{author}{\bibfnamefont{S.~P.}
  \bibnamefont{Walborn}}, \bibnamefont{and}
  \bibinfo{author}{\bibfnamefont{P.~H.~S.} \bibnamefont{Ribeiro}},
  \bibinfo{journal}{Phys. Rev. A} \textbf{\bibinfo{volume}{90}},
  \bibinfo{pages}{052118} (\bibinfo{year}{2014}),
  \urlprefix\url{https://link.aps.org/doi/10.1103/PhysRevA.90.052118}.

\bibitem[{\citenamefont{Kawabata et~al.}(2017)\citenamefont{Kawabata, Ashida,
  and Ueda}}]{Kawabata2017}
\bibinfo{author}{\bibfnamefont{K.}~\bibnamefont{Kawabata}},
  \bibinfo{author}{\bibfnamefont{Y.}~\bibnamefont{Ashida}}, \bibnamefont{and}
  \bibinfo{author}{\bibfnamefont{M.}~\bibnamefont{Ueda}},
  \bibinfo{journal}{Phys. Rev. Lett.} \textbf{\bibinfo{volume}{119}},
  \bibinfo{pages}{190401} (\bibinfo{year}{2017}),
  \urlprefix\url{https://link.aps.org/doi/10.1103/PhysRevLett.119.190401}.

\bibitem[{\citenamefont{Xiao et~al.}(2019)\citenamefont{Xiao, Wang, Zhan, Bian,
  Kawabata, Ueda, Yi, and Xue}}]{LXiao2019}
\bibinfo{author}{\bibfnamefont{L.}~\bibnamefont{Xiao}},
  \bibinfo{author}{\bibfnamefont{K.}~\bibnamefont{Wang}},
  \bibinfo{author}{\bibfnamefont{X.}~\bibnamefont{Zhan}},
  \bibinfo{author}{\bibfnamefont{Z.}~\bibnamefont{Bian}},
  \bibinfo{author}{\bibfnamefont{K.}~\bibnamefont{Kawabata}},
  \bibinfo{author}{\bibfnamefont{M.}~\bibnamefont{Ueda}},
  \bibinfo{author}{\bibfnamefont{W.}~\bibnamefont{Yi}}, \bibnamefont{and}
  \bibinfo{author}{\bibfnamefont{P.}~\bibnamefont{Xue}},
  \bibinfo{journal}{Phys. Rev. Lett.} \textbf{\bibinfo{volume}{123}},
  \bibinfo{pages}{230401} (\bibinfo{year}{2019}),
  \urlprefix\url{https://link.aps.org/doi/10.1103/PhysRevLett.123.230401}.

\bibitem[{\citenamefont{Bian et~al.}(2020)\citenamefont{Bian, Xiao, Wang,
  Onanga, Ruzicka, Yi, Joglekar, and Xue}}]{ZBian2020}
\bibinfo{author}{\bibfnamefont{Z.}~\bibnamefont{Bian}},
  \bibinfo{author}{\bibfnamefont{L.}~\bibnamefont{Xiao}},
  \bibinfo{author}{\bibfnamefont{K.}~\bibnamefont{Wang}},
  \bibinfo{author}{\bibfnamefont{F.~A.} \bibnamefont{Onanga}},
  \bibinfo{author}{\bibfnamefont{F.}~\bibnamefont{Ruzicka}},
  \bibinfo{author}{\bibfnamefont{W.}~\bibnamefont{Yi}},
  \bibinfo{author}{\bibfnamefont{Y.~N.} \bibnamefont{Joglekar}},
  \bibnamefont{and} \bibinfo{author}{\bibfnamefont{P.}~\bibnamefont{Xue}},
  \bibinfo{journal}{Phys. Rev. A} \textbf{\bibinfo{volume}{102}},
  \bibinfo{pages}{030201} (\bibinfo{year}{2020}),
  \urlprefix\url{https://link.aps.org/doi/10.1103/PhysRevA.102.030201}.

\end{thebibliography}
\bibliographystyle{apsrev}

\end{document}